\documentclass[12pt, onecolumn, draftclsnofoot]{IEEEtran}%

\usepackage{setspace} 

\usepackage[T1]{fontenc}
\usepackage{amsmath,amssymb,amsfonts,mathrsfs,bm}
\usepackage{amsthm}
\usepackage{cite}
\usepackage{array}
\usepackage[shortlabels]{enumitem}
\usepackage{graphicx}
\usepackage{url}
\usepackage{color}
\usepackage{algorithm,algorithmic}
\usepackage{multirow}
\usepackage{breqn}
\usepackage[table]{xcolor}
\usepackage[normalem]{ulem}
\usepackage{array}
\newcolumntype{L}[1]{>{\raggedright\let\newline\\\arraybackslash\hspace{0pt}}m{#1}}
\newcolumntype{C}[1]{>{\centering\let\newline\\\arraybackslash\hspace{0pt}}m{#1}}
\newcolumntype{R}[1]{>{\raggedleft\let\newline\\\arraybackslash\hspace{0pt}}m{#1}}
\usepackage{subfigure}

\usepackage{xparse}

\ifx\notloadhyperref\undefined
	\ifx\loadbibentry\undefined
		\usepackage[hidelinks]{hyperref} 
	\else
		\usepackage{bibentry}
		\makeatletter\let\saved@bibitem\@bibitem\makeatother
		\usepackage[hidelinks]{hyperref}
		\makeatletter\let\@bibitem\saved@bibitem\makeatother
	\fi
\else
\relax
\fi

\usepackage[capitalize]{cleveref}
\crefname{equation}{}{}
\Crefname{equation}{}{}
\crefname{claim}{claim}{claims}
\crefname{step}{step}{steps}
\crefname{line}{line}{lines}
\crefname{Theorem}{Theorem}{Theorems}
\crefname{Corollary}{Corollary}{Corollaries}
\crefname{Proposition}{Proposition}{Propositions}
\crefname{Lemma}{Lemma}{Lemmas}
\crefname{Definition}{Definition}{Definitions}
\crefname{Example}{Example}{Examples}
\crefname{Assumption}{Assumption}{Assumptions}
\crefname{Remark}{Remark}{Remarks}
\crefname{Theorem_A}{Theorem}{Theorems}
\crefname{Corollary_A}{Corollary}{Corollaries}
\crefname{Proposition_A}{Proposition}{Propositions}
\crefname{Lemma_A}{Lemma}{Lemmas}
\crefname{Definition_A}{Definition}{Definitions}

\ifx\useTheoremCounter\undefined
\newtheorem{Theorem}{Theorem}
\newtheorem{Corollary}{Corollary}
\newtheorem{Proposition}{Proposition}

\else
\newtheorem{Theorem}{Theorem}
\newtheorem{Corollary}[theorem]{Corollary}
\newtheorem{Proposition}[theorem]{Proposition}
\fi

\newtheorem{Definition}{Definition}
\newtheorem{Example}{Example}

\newtheorem{Corollary_A}{Corollary}[section]

\newtheorem{Lemma_A}{Lemma}[section]

\theoremstyle{remark}

\DeclareSymbolFont{bsfletters}{OT1}{cmss}{bx}{n}
\DeclareSymbolFont{ssfletters}{OT1}{cmss}{m}{n}
\DeclareMathSymbol{\bsfGamma}{0}{bsfletters}{'000}
\DeclareMathSymbol{\ssfGamma}{0}{ssfletters}{'000}
\DeclareMathSymbol{\bsfDelta}{0}{bsfletters}{'001}
\DeclareMathSymbol{\ssfDelta}{0}{ssfletters}{'001}
\DeclareMathSymbol{\bsfTheta}{0}{bsfletters}{'002}
\DeclareMathSymbol{\ssfTheta}{0}{ssfletters}{'002}
\DeclareMathSymbol{\bsfLambda}{0}{bsfletters}{'003}
\DeclareMathSymbol{\ssfLambda}{0}{ssfletters}{'003}
\DeclareMathSymbol{\bsfXi}{0}{bsfletters}{'004}
\DeclareMathSymbol{\ssfXi}{0}{ssfletters}{'004}
\DeclareMathSymbol{\bsfPi}{0}{bsfletters}{'005}
\DeclareMathSymbol{\ssfPi}{0}{ssfletters}{'005}
\DeclareMathSymbol{\bsfSigma}{0}{bsfletters}{'006}
\DeclareMathSymbol{\ssfSigma}{0}{ssfletters}{'006}
\DeclareMathSymbol{\bsfUpsilon}{0}{bsfletters}{'007}
\DeclareMathSymbol{\ssfUpsilon}{0}{ssfletters}{'007}
\DeclareMathSymbol{\bsfPhi}{0}{bsfletters}{'010}
\DeclareMathSymbol{\ssfPhi}{0}{ssfletters}{'010}
\DeclareMathSymbol{\bsfPsi}{0}{bsfletters}{'011}
\DeclareMathSymbol{\ssfPsi}{0}{ssfletters}{'011}
\DeclareMathSymbol{\bsfOmega}{0}{bsfletters}{'012}
\DeclareMathSymbol{\ssfOmega}{0}{ssfletters}{'012}

\DeclareMathOperator*{\argmax}{arg\,max}
\DeclareMathOperator*{\argmin}{arg\,min}

\newcommand{\qednew}{\nobreak \ifvmode \relax \else
      \ifdim\lastskip<1.5em \hskip-\lastskip
      \hskip1.5em plus0em minus0.5em \fi \nobreak
      \vrule height0.75em width0.5em depth0.25em\fi}

\newcommand{\norm}[1]{{\left\lVert{#1}\right\rVert}}

\newcommand{\cond}[2]{\left. {#1}\, \middle| \, {#2} \right.}

\DeclareDocumentCommand \P { g d() g } {%
	\IfNoValueTF {#3} 
	{%
		\IfNoValueTF {#1} 
		{%
			\IfNoValueTF {#2}
			{%
				\mathbb{P}%
			}%
			{%
				\mathbb{P}\left({#2}\right)%
			}%
		}%
		{%
			\IfNoValueTF {#2}
			{%
				\mathbb{P}_{#1}%
			}%
			{%
				\mathbb{P}_{#1}\left({#2}\right)%
			}%
		}%
	}%
	{%
		\IfNoValueTF {#1} 
		{%
			\mathbb{P}\left(\cond{#2}{#3}\right)%
		}%
		{%
			\mathbb{P}_{#1}\left(\cond{#2}{#3}\right)%
		}%
	}%
}

\DeclareDocumentCommand \E { g o g } {%
	\IfNoValueTF {#3} 
	{%
		\IfNoValueTF {#1} 
		{%
			\IfNoValueTF {#2}
			{%
				\mathbb{E}%
			}%
			{%
				\mathbb{E}\left[{#2}\right]%
			}%
		}%
		{%
			\IfNoValueTF {#2}
			{%
				\mathbb{E}_{#1}%
			}%
			{%
				\mathbb{E}_{#1}\left[{#2}\right]%
			}%
		}%
	}%
	{%
		\IfNoValueTF {#1} 
		{%
			\mathbb{E}\left[\cond{#2}{#3}\right]%
		}%
		{%
			\mathbb{E}_{#1}\left[\cond{#2}{#3}\right]%
		}%
	}%
}

\definecolor{gray90}{gray}{0.9}

\newcommand{\msout}[1]{\text{\color{green} \sout{\ensuremath{#1}}}}
\newcommand{\del}[1]{{\color{green}\ifmmode \msout{#1}\else\sout{#1}\fi}}

\newcommand{\hide}[1]{}

\newcommand{\figref}[1]{\figurename~\ref{#1}}
\renewcommand{\figurename}{Fig.}
\graphicspath{{./Figures/}} 
\usepackage{graphicx}
\usepackage{tikz}
\usepackage{color}
\usepackage{pgfplots}
\pgfplotsset{compat=1.5}

\providecommand{\U}[1]{\protect\rule{.1in}{.1in}}

\theoremstyle{definition}

\begin{document}
\title {On the Properties of Gromov Matrices and Their Applications in Network Inference}
\author{
    Feng~Ji, Wenchang~Tang, and Wee~Peng~Tay,~\IEEEmembership{Senior Member,~IEEE}%
 \thanks{This research is supported by the Singapore Ministry of Education Academic Research Fund Tier 1 grant 2017-T1-001-059 (RG20/17).}%
\thanks{F.~Ji, W.~Tang and W.~P.~Tay are with the School of Electrical and Electronic Engineering, Nanyang Technological University, 639798, Singapore (e-mail: jifeng@ntu.edu.sg, E150012@ntu.edu.sg, wptay@ntu.edu.sg).}
}

\maketitle

\begin{abstract}
The spanning tree heuristic is a commonly adopted procedure in network inference and estimation. It allows one to generalize an inference method developed for trees, which is usually based on a statistically rigorous approach, to a heuristic procedure for general graphs by (usually randomly) choosing a spanning tree in the graph to apply the approach developed for trees. However, there are an intractable number of spanning trees in a dense graph. In this paper, we represent a weighted tree with a matrix, which we call a Gromov matrix. We propose a method that constructs a family of Gromov matrices using convex combinations, which can be used for inference and estimation instead of a randomly selected spanning tree. This procedure increases the size of the candidate set and hence enhances the performance of the classical spanning tree heuristic. On the other hand, our new scheme is based on simple algebraic constructions using matrices, and hence is still computationally tractable. We discuss some applications on network inference and estimation to demonstrate the usefulness of the proposed method. 
\end{abstract}
\begin{IEEEkeywords}
	Gromov matrix, spanning tree, network inference and estimation 
\end{IEEEkeywords}

\section{Introduction}
Information is being propagated across networks like online social networks with increasing speed due to better connectivity. There has been recent increased interest in inference and estimation problems on networks. For example, information dynamics and social learning have been investigated in \cite{Gui2013, Kempe2003, Java2006, Leskovec2007, Tay:J12, SohTayQue:J13, Tay:J15, HoTayQue:J15}. Finding the sources of an infection or rumor diffusing in a network has been investigated in various works, including \cite{Shah2011, Pinto2012, Zhu2013, LuoTayLen14, JiTayVar:J17, Dong2013, LuoTay:C13, Lokho2013, LuoTayLen:J16}. Networks are usually modeled mathematically using graphs. Many research works (for example, \cite{Shah2011, Pinto2012, Zhu2013, LuoTayLen14, JiTayVar:J17}) on network inference adopts the following strategy: 
\begin{enumerate}[(i)]
	\item One first establishes theoretical results on tree networks. In many practical applications, it is reasonable to assume that information propagation follows a spanning tree. Furthermore, the uniqueness of a simple path between any pair of nodes greatly simplifies the theoretical analysis, making it tractable to obtain theoretically rigorous results.
	\item \label{it:atba} A tree-based algorithm or estimator is proposed based on the theoretical studies on trees.
	\item The algorithm or estimator is extended to general graphs using the spanning tree heuristic: a spanning tree of a selected set of nodes is found. The tree-based algorithm or estimator in \ref{it:atba} is then applied to this tree. An additional optimization procedure may be used to arrive at the final estimate for the general graph.  
\end{enumerate}
An example is given as follows.

\begin{Example}\label{example:infection}
	Consider the network infection source estimation problem in which an infection starting from a source node propagates along the edges of a graph $G = (V,E)$. A snapshot $U\subset V$ of all the infected nodes is observed, and we wish to estimate the identity of the source node (see for example \cite{Shah2011,Tan2016,JiTayVar:J17}). One possible approach is to maximize a real-valued estimator function $e(s,T)$ over each candidate source node $s \in V$, and each tree $T$ rooted at $s$ that spans the infected nodes $U$. To make this optimization procedure computationally tractable for a general graph, for each candidate source node $s$, a random  breadth-first search (BFS) tree $T_{\mathrm{BFS}}(s)$ is usually chosen (e.g., \cite{Shah2011,Tan2016,JiTayVar:J17}), and the source node is estimated by $\argmax_{s\in V} e(s, T_{\mathrm{BFS}}(s))$.  
	
	The idea behind this approach is that for each realization of the infection process, the infected set $U$ is likely to be spanned by a BFS tree rooted at the true source node. The BFS tree is unknown \emph{a priori}. By randomly choosing a BFS tree for each $s$, even if $s$ happens to be the true source node, it is unlikely that the chosen BFS tree is the actual infection tree if $G$ is a dense graph. An arguably better approach is to find $\max_{s, T\in \mathcal{T}_s} e(s, T)$, where $\mathcal{T}_s$ is a family of trees rooted at $s$ that span $U$, and which is restricted to be small enough so that the maximization procedure is computationally tractable.    	
	
	There are other variants of this problem. For example, instead of observing a snapshot status of all the nodes of $G$, one may observe the infection timestamps of a small portion of the network. The source is estimated using such timestamp information (e.g., \cite{Pinto2012, Louni2014, TanTay:C17, Tan18}). Similar heuristics involving BFS trees have also been proposed.   
\end{Example}

The family of trees $\mathcal{T}_s$ rooted at $s$ in \cref{example:infection} needs to be chosen carefully. For example, in the extreme case, Cayley's formula states that the number of spanning trees in a complete graph with $n$ vertices is $n^{n-2}.$ In this paper, our aim is to propose an approach to construct a reasonable $\mathcal{T}_s$ to partially overcome such an intractability problem. We propose the use of Gromov matrices (which we define in Section~\ref{sec:cmow}) to represent a weighted tree in a graph. Then by taking appropriate convex combinations of a small set of such Gromov matrices, we can generate a tractable convex set $\mathcal{T}_s$ over which our estimator function can be optimized.    

This paper takes its root in \cite{Tan18}, in which we introduced the use of a Gromov matrix to represent an \emph{unweighted} tree. In this paper, we further refine and generalize our Gromov matrix approach to weighted graphs, prove various properties of the Gromov matrices and demonstrate how to apply Gromov matrices in several network inference problems. A preliminary version of this paper was published in the conference \cite{JiTanTay18}.

The rest of the paper is organized as follows. In Section~\ref{sec:cmow}, we introduce the notion of a Gromov matrix and its properties. In Section~\ref{sec:ico}, we demonstrate geometrically how a tree is constructed from a single point by a series of simple steps. The discussions lead to an iterative way to construct a Gromov matrix, and provides tools to investigate various properties of such matrices. In Section~\ref{sec:para}, we explain how to construct a parametric family of trees from a fixed finite set of trees using Gromov matrices. We discuss properties of this construction. We give applications of our method on network inference and estimation in Section~\ref{sec:app}, and conclude in Section~\ref{sec:con}. Miscellaneous discussions and technical proofs of some results are given in the appendices.

\emph{Notations:} Given a matrix $M$, we use $M(i,j)$ to denote its $(i,j)$-th entry. We write $M^{-1}$ for its inverse. For a real symmetric matrix $M$, we denote its smallest eigenvalue by $\lambda_{\min}(M)$. For two nodes $u$ and $v$ in a tree, we use $[u,v]$ to denote the path from $u$ to $v$ and $(u,v]$ to denote the path with the end node $u$ excluded. Other notations like $[u,v)$ and $(u,v)$ are defined similarly. For a graph $G$, let $d_G(u,v)$ be the sum of the edge weights of the shortest path $P$ in $G$ between vertices $u$ and $v$. 

\section{Gromov Matrices of Weighted Trees and the Three-point Condition} \label{sec:cmow}

Suppose that $T$ is a weighted, undirected tree. The weight of each edge $(u,v)$ is a positive real number. 

\begin{Definition} \label{defn:fabp}
	Let $s$ in $T$ be a fixed node. For $u,v$ in $T$, their \emph{Gromov product} (see Definition 2.6 of \cite{Kapovich2002}) w.r.t. $s$ is defined as 
	\begin{align}
	(u,v)_{s} = \frac{1}{2}(d_T(u,s)+d_T(v,s)-d_T(u,v)).
	\end{align}
	For an ordered set of vertices $V=\{v_1,\ldots,v_n\}\subset T$ that does not contain $s$, the associated \emph{Gromov matrix} $M$ is the matrix with $(i,j)$-th entry $M(i,j)=(v_i,v_j)_s$. We say that $M$ has base $(T,s,V)$, where $T$ is known as its base tree, $s$ its base vertex, and $V$ its base set.
\end{Definition}

We view a weighted tree as a metric space. Two weighted trees $T_1$ and $T_2$ are said to be \emph{isometric} to each other if there is a bijection $\varphi: T_1 \to T_2$ such that $d_{T_1}(u,v) = d_{T_2}(\varphi(u),\varphi(v))$ for any $u,v \in T_1$. The map $\varphi$ is called an isometry. We say that two basis $(T_1,s_1,V_1)$ and $(T_2,s_2,V_2)$ are \emph{isometrically equivalent} to each other if there is an isometry $\varphi: T_1 \to T_2$ such that $\varphi(s_1)=s_2$ and $\varphi(V_1)= V_2$. 

Suppose for a base $(T,s,V)$, $T$ spans $V\cup \{s\}$. Proposition 1 of \cite{Tan18} may be generalized (using a similar proof) as follows: for each Gromov matrix, its base is uniquely determined, up to isometric equivalence. This observation tells us that the Gromov matrix contains all the information we want to describe a base.

This one to one correspondence shows that the Gromov matrix can be useful as an algebraic tool to study trees. The notion of Gromov product is used extensively in the theory of $\delta$-hyperbolic spaces. In particular, it is used to define boundary (points at infinity) of a hyperbolic space. Intuitively, if two nodes have a large Gromov product w.r.t. a base vertex, then they are geometrically similar if we look at a neighborhood of the base vertex bounded by the Gromov product. In this paper, we use this perspective to understand trees using the algebraic tool of Gromov matrices.

In Corollary~\ref{coro:ivin} in the sequel, we show that the Gromov matrix $M$ is positive definite and symmetric. Conversely, not all positive definite symmetric matrices are Gromov matrices of a (weighted) tree as above, even if we assume the entries are non-negative numbers. Simple examples include: 
\begin{align*}
\begin{bmatrix}
1 & 3 \\
3 & 10 	
\end{bmatrix},
\end{align*}
where the top-left diagonal entry $1$ is smaller than the off-diagonals; and 
\begin{align*}
\begin{bmatrix}
3 & 2 & 1\\
2 & 3 &	2\\
1 & 2 & 3
\end{bmatrix},
\end{align*}
where the top-right and bottom-left entries are too small. We give below a condition that guarantees a matrix being the Gromov matrix of a weighted tree.

\begin{figure}[!t] 
	\centering
	\includegraphics[width=0.45\columnwidth]{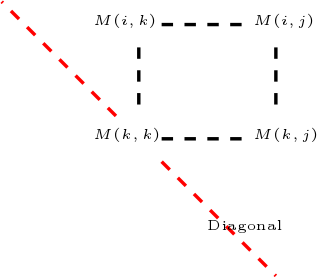}
	\caption{Illustration of the notion of corners.} \label{fig:ultra}
\end{figure}

\begin{Definition} \label{defn:lmba}
	Let $M$ be an $n\times n$ matrix. We say that $M$ satisfies the \emph{three-point condition} if the following holds: for any distinct indices $i,j,k$ and any permutation $x,y,z$ of $\{M(i,j), M(i,k), M(k,j)\}$, we have $x \geq \min\{y,z\}$ and equality holds if $y\neq z.$
	
	For convenience, the triple $x,y,z$ are called \emph{corners} of the same rectangle (see Figure~\ref{fig:ultra}).
\end{Definition}

The three-point condition\footnote{The name is due to its resemblance to the four-point condition in the theory of $\delta$-hyperbolic metric space (cf.\ remark after Proposition 6.13 of \cite{Bow06}). In our case, the base $s$ is fixed, therefore we only have to concern about the three remaining points.} requires that $\min\{x,y,z\}$ must occur at least twice among $\{x,y,z\}$, i.e., the two smallest values among the three must be equal. 

\begin{figure}[!htb] 
	\centering
	\includegraphics[width=0.62\columnwidth]{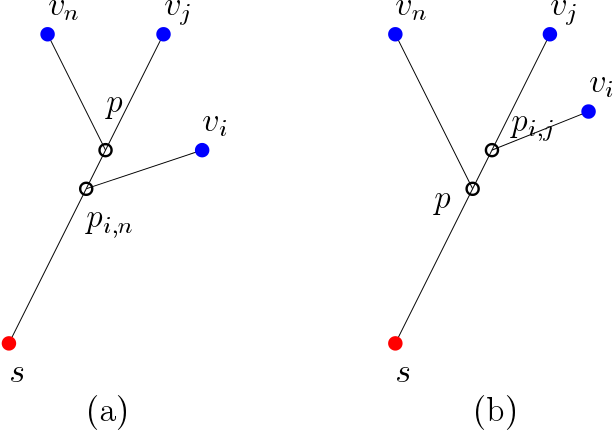}
	\caption{Illustration of the proof of \cref{thm:ansm}.} \label{fig:15}
\end{figure}

\begin{Theorem} \label{thm:ansm}
	An $n\times n$ symmetric matrix $M$ is the Gromov matrix of a weighted tree if and only if the following conditions hold for $M$: 
	\begin{enumerate}[(a)]
		\item\label{ansm_pos} $M$ has non-negative entries and all the diagonal entries of $M$ are positive.
		\item\label{ansm_diag} For all $1\leq i,j\leq n$, $M(i,i)\geq M(i,j)$.
		\item\label{ansm_three-point} $M$ satisfies the three-point condition.
	\end{enumerate}	
\end{Theorem}
\begin{IEEEproof}
	Suppose that $M$ is a Gromov matrix with base $(T,s,V)$. It is easy to show that conditions \ref{ansm_pos} and \ref{ansm_diag} hold from \cref{defn:lmba}. To verify condition \ref{ansm_three-point}, suppose that $v_i,v_j,v_k\in V$ correspond to the distinct indices $i,j,k$ of the matrix $M$. The branch point $p_{i,j}$ is defined as the unique vertex of $T$ such that $[s,p_{i,j}]=[s,v_i]\cap [s,v_j]$. The branch points $p_{i,k}$ and $p_{k,j}$ are defined similarly. If $M(i,k)> M(k,j)$, then $p_{k,j} \in [s, p_{i,k})$. Hence, we must have $p_{i,j}= p_{k,j}$ and $M(i,j)=M(k,j)$. The case $M(i,k)<M(k,j)$ is similar. If $M(i,k)=M(k,j)$, then $p_{i,k}=p_{k,j}\in [s,p_{i,j}]$. Hence $M(i,j)\geq M(i,k)=M(k,j).$ A similar argument holds for the two inequalities $M(i,k)\geq \min\{M(i,j),M(k,j)\}$ and $M(k,j)\geq \min\{M(i,k),M(i,j)\}$.
	
	We prove the converse by induction on $n$. The case $n=0, 1,2$ are clearly true. Suppose the converse holds for all $k\times k$ symmetric matrices satisfying the conditions \ref{ansm_pos} -- \ref{ansm_three-point} and for all $k<n$. Consider a $n\times n$ symmetric matrix $M$ satisfying conditions \ref{ansm_pos} -- \ref{ansm_three-point}. By the induction hypothesis, we can construct a base $(T', s, V')$ associated with the top-left $(n-1)\times (n-1)$ matrix of $M$. We add an additional node $v_n$ to $V'$ as follows: Let $1\leq j\leq n-1$ be the index such that $M(j,n)=\max\{M(i,n): 1\leq i\leq n-1\}$ with corresponding vertex $v_j\in V'$. Let $p$ be the vertex on the path $[s,v_j]$ such that $d_{T'}(s,p)=M(j,n)$; and if it does not exist, we just let introduce it as a new vertex. We add  an edge $(v_n,p)$ of weight $M(n,n)-M(j,n) > 0$ to $T'$ at $p$ and set $T =T'\cup \{v_n\}$ and $V= V'\cup \{v_n\}.$ The tree $T$ is a well-defined weighted tree due to condition \ref{ansm_diag}. Furthermore, $M(j,n)=(v_j,v_n)_s$ by construction.
	
	To see that $(T,s,V)$ is a base of $M$, consider any index $i\neq j$ such that $1\leq i <n$ with corresponding vertex $v_i$. From the induction hypothesis, it suffices to show that $M(i,n)=(v_i,v_n)_s$ in $T$. We consider two cases below (see \figref{fig:15} for illustrations).
	
	Suppose that $M(i,n) < M(j,n)$. Then, from the three-point condition \ref{ansm_three-point}, we have $M(i,j)=M(i,n)$. By our construction of $p$, the branch point $p_{i,n}$ is on the path $[s,p]$. This implies that $(v_i,v_n)_s=(v_i,v_j)_s=M(i,j)=M(i,n)$, where the penultimate equality is due to the induction hypothesis.
	
	On the other hand, if $M(i,n)=M(j,n)$, then from the three-point condition \ref{ansm_three-point}, we have $M(i,n)=M(j,n) \leq M(i,j)$. Therefore, in our construction of $p$, we have $p \in [s,p_{i,j}]$ and $(v_i,v_n)_s = (v_j,v_n)_s = M(j,n) = M(i,n)$, where the penultimate equality is due to our construction. The converse is now proved, and so is the theorem.  
\end{IEEEproof}

\begin{figure}[!htb] 
	\centering
	\includegraphics[width=0.85\columnwidth]{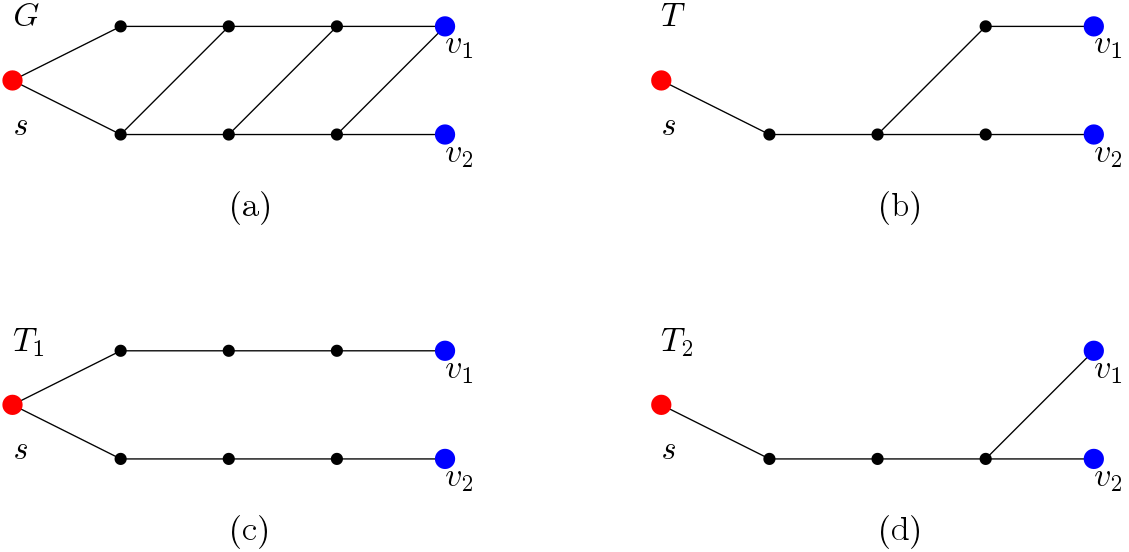}
	\caption{Let $G$ be the graph in (a) and $V=\{v_1,v_2\}$. Consider three spanning trees $T, T_1, T_2$ of $V$ and the base vertex $s$, as shown in (b), (c) and (d) respectively. Suppose $M_T, M_{T_1}$ and $M_{T_2}$ are the corresponding associated Gromov matrices. Then $M_T=1/3M_{T_1}+2/3M_{T_2}.$} \label{fig:4}
\end{figure}

We have the following immediate corollary regarding convex combinations of special Gromov matrices (see \figref{fig:4} for an example).

\begin{Corollary} \label{coro:lmam}
	Suppose that $M_1$ and $M_2$ are Gromov matrices of weighted trees and $M_2$ is diagonal. Then for any $\theta\in [0,1]$, the convex combination $M=\theta M_1 + (1-\theta) M_2$ is the Gromov matrix of a weighted tree.	
\end{Corollary}
\begin{IEEEproof}
	It is easy to verify that $M$ satisfies the three conditions of Theorem~\ref{thm:ansm}.
\end{IEEEproof}

The conclusion does not hold in general if both $M_1$ and $M_2$ are not diagonal. In Section~\ref{sec:para} we describe a procedure to modify the convex combination of multiple Gromov matrices to a Gromov matrix.

We end this section by describing a few simple special examples to illustrate some basic properties of the Gromov product. 
\begin{Example}
	Let $s$ be a fixed base vertex of the tree $T$ and $u,v$ are two distinct nodes of $T$. 
	\begin{enumerate}[(a)]
		\item If the Gromov product $(u,v)_s$ is $0$, then $s$ is on the unique path connecting $u$ and $v$. Therefore, if $M_T$ is a diagonal matrix, then $T$ is a star centered at $s$.
		\item The Gromov products $(u,v)_s = (u,u)_s$ if and only if $u$ is on the unique path connecting $s$ and $v$. 
		\item The span of $\{s,u,v\}$ is either a line or star with $3$ leaves. The latter case happens if and only if $0<(u,v)_s < \min\{(u,u)_s,(v,v)_s\}.$ 
	\end{enumerate}
\end{Example}

\section{Iterative Construction of a Gromov Matrix and Some Spectral Properties} \label{sec:ico}

In this section, we show positive definiteness of the Gromov matrix $M$ of a weighted tree $T$ as a byproduct of finding a lower bound of its least singular value, i.e., the smallest eigenvalue of $M$. To do so, we describe an iterative way to construct $M$ from the empty matrix (of size $0$); and each step in the algebraic construction has a corresponding geometric counterpart, revealing additional geometric properties of $T$. We delegate some long technical proofs to the appendices.


For two square matrices $M$ and $N$ of sizes $k$ and $l$ respectively, we introduce the following operations, called the \emph{Gromovication operations} for convenience:
\begin{enumerate}[(a)]
	\item Initialization: for $a>0$, replace the empty matrix $M$ by the $1\times 1$-matrix $M=(a)$.
	\item Direct sum: $M\oplus N$ is the $(k+l)\times (k+l)$ square matrix with $M$ and $N$ forming the diagonal blocks; and $0$ elsewhere. 
	\item Extension I (of $M$): for an integer $a>0$, $\phi_a(M)$ is the $k\times k$ matrix obtained from $M$ by adding $a$ to each entry.
	\item Extension II (of $M$): for two positive integers $a,b$ such that $a\geq b> 0$, $\phi_{a,b}(M)$ is the $(k+1)\times (k+1)$ matrix obtained from $M$ by adding $a$ to each entry of $M$, followed by increasing the number of rows and columns of $M$ each by one, and setting $b$ as the entries of the $(k+1)$-th row and column.
\end{enumerate}

\begin{Proposition} \label{prop:mit}
	$M$ is the Gromov matrix of a weighted tree with vertex set $V\neq \emptyset$ if and only if for some permutation matrix $P$, $M'=PMP^{-1}$ can be obtained from the empty matrix by a finite sequence of Gromovication operations, starting with an initialization step.
\end{Proposition}
\begin{IEEEproof}
	See \cref{proof:prop:mit}.
\end{IEEEproof}

Suppose that $M$ is a Gromov matrix. We study the its least singular value, which is defined as $\lambda_{\min}(M)$, the minimal eigenvalue of $M$. As a corollary, we prove that $M$ is always positive definite.

\begin{Proposition} \label{prop:wht}
	Let $M$ and $N$ be Gromov matrices. We have the following results regarding the Gromovication operations. 
	\begin{enumerate}[(a)]
		\item\label{prop:wht:a} Direct sum: 
		\begin{align*}
		\lambda_{\min}(M\oplus N)=\min\{\lambda_{\min}(M), \lambda_{\min}(N)\}.
		\end{align*}
		\item Extension I: For any positive integer $a$,
		\begin{align*}
		\lambda_{\min}(\phi_a(M))\geq \lambda_{\min}(M). 
		\end{align*}
		\item Extension II: For any integers $a \geq b >0$,
		\begin{enumerate}[(a)]
			\item If $a>b$,
			\begin{align*}
			\lambda_{\min}(\phi_{a,b}(M))\geq \min\{\lambda_{\min}(M),b-b^2/a\}.
			\end{align*}
			\item If $a=b$,
			\begin{align*}
			\lambda_{\min}(\phi_{a,a}(M))\geq \frac{\lambda_{\min}(M)}{n+1+\lambda_{\min}(M)/a},
			\end{align*}
			where $n$ is the number of rows of $M$.
		\end{enumerate}		
	\end{enumerate}
\end{Proposition}
\begin{IEEEproof}
	See \cref{proof:prop:wht}.
\end{IEEEproof}

\begin{figure}[!t] 
	\centering
	\includegraphics[width=0.4\columnwidth]{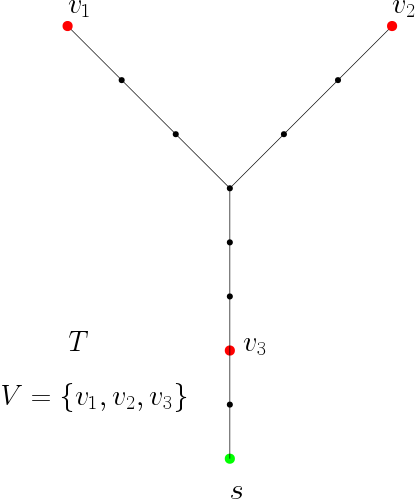}
	\caption{Suppose $M$ is a Gromov matrix with base $(T,s,V)$, where $T$ is shown in the figure, $V$ consists of the three red vertices; and $s$ is the green vertex. Using our lower bound for direct sum and extension II, we see immediately that $\lambda_{\min}(M_T)\geq \min\{\min\{3,3\},2-2^2/5\}=1.2$. One can also compute that the eigenvalues (in increasing order) of $M$ are $13.69,3,1.315$; and the lower bound $1.2$ is quite close in this case.} \label{fig:3}
\end{figure}

Proposition~\ref{prop:wht} can be useful in computation as shown in the example given in \figref{fig:3}. Another immediate consequence is the following result.

\begin{Corollary} \label{coro:ivin}
	Suppose that $M$ is a Gromov matrix with base $(T,s,V)$. If $V$ is non-empty, then $M$ is positive definite.	
\end{Corollary}
\begin{IEEEproof}
	From Proposition~\ref{prop:mit}, there exists a permutation matrix $P$ such that $M'=PMP^{-1}$ is formed by a finite sequence of Gromovication operations. Note that $M$ and $M'$ have the same eigenvalues, thus it suffices to show that $M'$ is positive definite. Since $V$ is non-empty, we start with an initialization, making the matrix positive definite. From Proposition~\ref{prop:wht}, the matrix $M$ remains positive definite for any additional Gromovication operations, and the corollary is proven.
\end{IEEEproof}

\begin{figure}[!htb] 
	\centering
	\includegraphics[width=0.75\columnwidth]{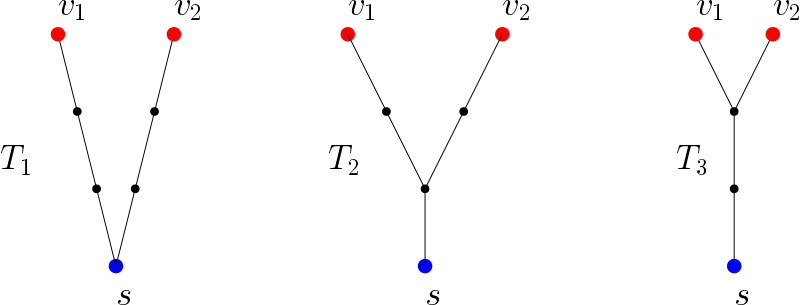}
	\caption{In the three trees $T_1,T_2,T_3$, it is easy to compute that the least singular value of the associated Gromov matrices $M_{T_1}, M_{T_2}, M_{T_3}$ (w.r.t. $s$ and $\{v_1,v_2\}$) are $3, 2, 3-\sqrt{5}$, respectively. Geometrically, $T_3$ is least branched off, while $T_1$ is most branched off.} \label{fig:14}
\end{figure}

Intuitively, the least singular value of a Gromov matrix $M$ encodes geometric information of its base tree $T$. If the least singular value of $M$ is small, the leaves of $T$ tend to be less branched off. We illustrate this intuition by a simple example in \figref{fig:14}.

\section{Convex Combinations of Gromov Matrices} \label{sec:para}

In this section, we discuss two different ways to construct new matrices from existing Gromov matrices based on convex combination. We also compare the advantages of each approach. Through a case study, we provide further insights into how well convex combinations of Gromov matrices can approximate an information propagation path in a network.

\subsection{Two Types of Convex Combinations}\label{subsec:convex}

Suppose we are given a finite collection of $n\times n$ Gromov matrices ${\bf M}=\{M_1,\ldots, M_k\}$, all of the same dimensions. For each $k$-tuple of numbers $\alpha=(\alpha_1,\ldots,\alpha_k)\in [0,1]^k$ such that $\sum_{i=1}^{k}\alpha_i=1$, define a new matrix as $M_{\alpha} = \sum_{i=1}^{k}\alpha_iM_i.$ We call $M_\alpha$ the \emph{convex combination} of ${\bf M}$, for convenience. 

To gain some geometric intuition, let $M_1$ be a $2\times 2$ Gromov matrix (such that the associated base is $(T, s, \{u,v\})$) and $M_2$ be the diagonal matrix formed by the diagonal of $M_1$. In $M_{\alpha}$, the off-diagonal entries of $M_1$ are being scaled down. Let $p$ be the joint point at the intersection of the paths $[s,u]$ and $[s,v]$. Taking convex combination decreases $[s,p]$, which means that geometrically, the joint point is being re-positioned along both the paths $[s,u]$ and $[s,v]$. Therefore, the convex combination can be used as a tool that re-scales the edge weights of a tree. 

In general, the convex combination $M_{\alpha}$ is not always a Gromov matrix of any weighted tree, as the three-point condition is not preserved under matrix addition. There are various ways to modify the entries of $M_{\alpha}$ to obtain a Gromov matrix. We provide one such procedure here, which turns out to be very natural. Consider the base case where $n=3$ and without loss of generality, we assume that $M_{\alpha}(1,2)\geq M_{\alpha}(1,3)> M_{\alpha}(2,3)$. To ensure \emph{the three-point condition and minimum changes to the ordering}, we essentially have 2 options: (1) replace $ M_{\alpha}(2,3)$ by $ M_{\alpha}(1,3)$; or (2) replace $ M_{\alpha}(1,3)$ by $ M_{\alpha}(2,3)$. For each option, we only need to make \emph{one change} in the top-right part of $M_{\alpha}$. For these two options, one increases the entries of $M_{\alpha}$ while the other decreases the entries. As we can scale down the matrix entries by taking convex combination with a diagonal matrix, our preferred choice is to increase the entries $M_{\alpha}$ in this procedure using option (1). We then apply option (1) whenever we encounter triples violating the three-point condition. A systematic way to do this is described in the following.

First set $M_{\alpha}'=M_{\alpha}$. Let ${\bf U}(M'_{\alpha})=(x_1,\ldots, x_{n(n-1)/2})$ denote the sequence of entries $M_{\alpha}'(i,j)$ with $i<j$ (i.e., the upper-triangular block excluding the diagonal) in non-increasing order with $x_1 \geq \ldots \geq x_{n(n-1)/2}$. If $x_t$ corresponds to $M_{\alpha}(i,j)$, then $(i,j)$ is called the matrix indices of $x_t$. For each $t=1,\ldots,n(n-2)/2$ in increasing order, we perform the following steps:
\begin{enumerate}[(a)]
	\item Search for every pair $(x_u, x_v)$ satisfying the following two conditions:
	\begin{itemize}
		\item \label{it:ua} $u<t<v$; and
		\item \label{it:xct} $x_u, x_t, x_v$ are corners of the same rectangle (Definition \ref{defn:lmba}). 
	\end{itemize}
	\item If the matrix indices of $x_v$ is $(i,j)$, we perform the following two steps:
	\begin{itemize}
		\item Set $x_v=x_t$ and $M'_{\alpha}(i,j)=x_t$.
		\item Move $x_v$ to be immediately after $x_t$ in ${\bf U}(M'_{\alpha})$. 
	\end{itemize}
	In doing so, the matrix indices of  $x_{t+1}$ becomes $(i,j)$ and the sequence ${\bf U}(M'_{\alpha})$ remains non-increasing.
\end{enumerate}
Once the above procedure is completed, we symmetrize $M_{\alpha}'$ by assigning $M_{\alpha}'(i,j)=M_{\alpha}'(j,i)$ for all $i>j$. We call $M_{\alpha}'$ a \emph{Gromovied convex (G-convex) combination} of ${\bf M}$ w.r.t.\ $\alpha$. Note that for each $t$, there are at most $n$ pairs $(x_u,x_v)$ satisfying condition (a). The computational complexity of forming $M_\alpha$ is $O(kn^2)$. Therefore, the complexity of forming a G-convex combination $M_\alpha'$ is $O(kn^2+n^3)$.

As we are enforcing the three-point condition each time we make an adjustment, the matrix $M_{\alpha}'$ is the Gromov matrix of a weighted tree by Theorem \ref{thm:ansm}. The matrix $M_{\alpha}'$ has the property that $M_{\alpha}'(i,j)\geq M_{\alpha}(i,j)$ for any $1\leq i,j\leq n.$ Moreover, if $M_{\alpha}$ is already a Gromov matrix, then $M_{\alpha}'$ is the same as $M_{\alpha}.$

Throughout the rest of this paper, when we talk about the convex or G-convex combination of a set of Gromov matrices ${\bf M}$, we assume for convenience that every Gromov matrix in ${\bf M}$ has the same base vertex and base set.

Let $G=(V_G,E)$ be a general graph, $U=\{u_1,\ldots, u_n\}\subset V_G$ be a fixed set of distinct vertices of $G$ and $s\in V_G$ a fixed node. In general, the spanning trees of $U$ rooted at $s$ are not unique and it is intractable to find all of them. However, if we are able to find a finite subset of spanning trees, we are able to use the G-convex combination construction to form a large family of trees $\mathcal{F}_s$. Doing so, we obtain a much larger candidate set than a randomly selected spanning tree, partially overcoming the intractability problem mentioned above. This can be useful in dealing with certain inference and estimation problems on graphs (see Section~\ref{sec:app}). Note however that although each G-convex combination has a corresponding tree, this tree need not be a subgraph of $G$. On the other hand, we hope that $\mathcal{F}_s$ captures some distinctive structural features of $s$ in $G$. We provide one such feature as follows.

\begin{Definition}\label{def:type}
	Let $T$ be a tree in a graph $G$ spanning a subset of distinct vertices $U=\{u_1,u_2,u_3\}$ and a fixed base vertex $s\notin U$. The \emph{type} of $U$ in $T$ is a partition of $U=U_1\cup U_2$ such that $U_1=\{u_i,u_j\}$ and $U_2=\{u_l\}$ (where $(i,j,l)$ is a permutation of the indices $1,2,3$) and $(u_i,u_l)_s= (u_j,u_l)_s\leq (u_i,u_j)_s$.
\end{Definition}

Some thought will convince the reader that Definition~\ref{def:type} is non-vacuous and the type of any set of three distinct vertices exists. Furthermore, the type of $U=\{u_1,u_2,u_3\}$ is not unique if and only if $(u_1,u_2)_s= (u_1,u_3)_s= (u_2,u_3)_s$. Intuitively, the notion of type describes the relative positions of nodes in a tree w.r.t.\ a fixed base node. We provide Example~\ref{eg:ltbt} below to illustrate this notion. Before that, we first prove an elementary result.   

Suppose we form the convex combination $M_{\alpha}=\theta M_1 + (1-\theta)M_2$, where $\alpha=(\theta,1-\theta)$ and $\theta \in [0,1]$, of two Gromov matrices $M_1$ and $M_2$, with $M_2$ being a diagonal matrix. Let the trees corresponding to $M_\alpha$, $M_1$ and $M_2$ be $T_{\alpha},T_1$, and $T_2$, respectively. For any $U=\{u_1,u_2,u_3\}$, its type in $T_{\alpha}$ is the same as its type in $T_1.$ This observation is not true in general if $M_2$ is not diagonal. Instead, we have the following weaker statement regarding the convex combination of an arbitrary finite set of Gromov matrices and type.

\begin{Proposition} \label{prop:ssia}
	Let $\{M_1,\ldots,M_k\}$ be a finite set of Gromov matrices with corresponding base trees $\{T_1,\ldots, T_k\}$ with the common base vertex $s$ and base set $V$. Consider $U=\{u_1,\ldots,u_n\} \subset V$ a subset of distinct vertices. For any convex combination weight vector $\alpha$, let $M_\alpha$ and $M'_\alpha$ be the convex and G-convex combination of $\{M_1,\ldots,M_k\}$, respectively. Suppose $T_{\alpha}'$ is the tree corresponding to $M_{\alpha}'$. For any triplet $\{u_i,u_j,u_l\}\subset U,$ if 
	\begin{align}\label{ineq:ssia}
	M_{\alpha}(i,j)= M_{\alpha}'(i,j)\geq M_{\alpha}'(i,l) = M_{\alpha}'(j,l),
	\end{align}
	then a type of $\{u_i,u_j,u_l\}$ in $T_{\alpha}'$ is the same as its type in one of $T_1,\ldots, T_k$.
\end{Proposition}
\begin{IEEEproof}
	From \eqref{ineq:ssia}, we can take $\{u_i,u_j\}\cup \{u_l\}$ as a type in $T_{\alpha}'$. As $M_{\alpha}(i,j)\geq M'_{\alpha}(i,l) = M'_{\alpha}(j,l)$ and each entry of $M'_{\alpha}$ is at least as large as the corresponding entry of $M_{\alpha}$, we have $M_{\alpha}(i,j)\geq M_{\alpha}(i,l)$ and $M_{\alpha}(i,j)\geq M_{\alpha}(l,j)$. 
	
	Suppose that no type of $\{u_i,u_j,u_l\}$ in $T_{\alpha}'$ is the same as its type in any of $\{T_1,\ldots,T_k\}$, i.e., for each $T_h\in \{T_1,\ldots, T_k\}$, $\{u_i,u_j\}\cup \{u_l\}$ is not a type in $T_h$. Therefore in $T_h$, we have either $(u_i,u_l)_{s}>(u_i,u_j)_{s}$ or $(u_j,u_l)_{s}>(u_i,u_j)_{s}$ from the three-point condition. Taking convex combination, we have either $M_{\alpha}(i,l)>M_{\alpha}(i,j)$ or $M_{\alpha}(l,j)>M_{\alpha}(i,j).$ This gives a contradiction. 
\end{IEEEproof}

Proposition~\ref{prop:ssia} places restrictions on $T_{\alpha}'$, the tree corresponding to a G-convex combination $M_\alpha'$. In particular, certain triplet node types must come from one of the components forming the G-convex combination, which corresponds to a spanning tree in the graph. Heuristically, taking G-convex combinations has two basic effects: (1) re-scaling distance and (2) mixing up types of the spanning trees in the components forming the G-convex combination. We illustrate the concepts and ideas of this section by the following example. 

\begin{Example} \label{eg:ltbt}
	Consider \figref{fig:1}. Let $T_1$ be the tree with Gromov matrix 
	$$
	M_1 = \begin{bmatrix}
	4 & 1 & 3 & 1 \\
	1 & 4 &	1 & 1 \\
	3 & 1 & 4 & 1 \\
	1 & 1 & 1 & 4
	\end{bmatrix},
	$$ 
	and $T_2$ be the tree with Gromov matrix 
	$$
	M_2 = \begin{bmatrix}
	4 & 1 & 1 & 1 \\
	1 & 4 &	3 & 2 \\
	1 & 3 & 4 & 2 \\
	1 & 2 & 2 & 4
	\end{bmatrix},
	$$
	both of which has base set $\{u_1,\ldots,u_4\}$. In $T_1$, $\{u_1,u_3,u_4\}$ is of type $\{u_1,u_3\}\cup \{u_4\}$, while in $T_2$, $\{u_1,u_3,u_4\}$ is of type $\{u_3,u_4\}\cup \{u_1\}.$ Hence $\{u_1,u_3,u_4\}$ are of different types in $T_1$ and $T_2$. Let $\alpha=(1/2,1/2)$, with the convex combination of $M_1$ and $M_2$ given by 
	$$
	M_{\alpha} =
	\begin{bmatrix}
	4 & 1 & 2 & 1 \\
	1 & 4 &	2 & 1.5 \\
	2 & 2 & 4 & 1.5 \\
	1 & 1.5 & 1.5 & 4
	\end{bmatrix},
	$$ 
	which is not a Gromov matrix. On the other hand, the G-convex combination matrix is 
	$$
	M_{\alpha}'=
	\begin{bmatrix}
	4 & 2 & 2 & 1.5 \\
	2 & 4 &	2 & 1.5 \\
	2 & 2 & 4 & 1.5 \\
	1.5 & 1.5 & 1.5 & 4
	\end{bmatrix}.$$ 
	The tree $T_{\alpha}'$ is shown in \figref{fig:1}. In $T_{\alpha}$, $\{u_1,u_3,u_4\}$ is of type $\{u_1,u_3\}\cup \{u_4\}$, the same as its type in $T_1$.
	
	\begin{figure}[!htb] 
		\centering
		\includegraphics[width=0.85\columnwidth]{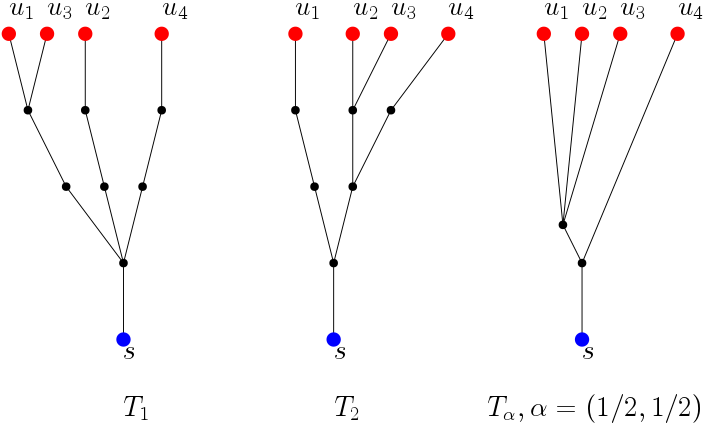}
		\caption{Illustration of $T_1, T_2$ and $T_{\alpha}$ in Example \ref{eg:ltbt}.} \label{fig:1}
	\end{figure}
\end{Example}

The set of symmetric matrices of fixed size is a vector space. Given two fixed Gromov matrices $M_1, M_2$ of the same size $n$, for $\theta$ varying from $0$ to $1$, the  convex combinations $M_{\alpha}$, where $\alpha=(\theta,1-\theta)$, give a linear path from $M_1$ to $M_2$ in the convex cone of positive definite symmetric matrices of size $n$. However, this linear path may not stay in the space of Gromov matrices. On the other hand, the family of G-convex combination matrices $M_\alpha'$ traces a continuous path between the end points $M_1$ and $M_2$, as shown in the following theorem. A path is \emph{piecewise linear} if it is continuous and is a finite union of line segments of the vector space.

\begin{Theorem} \label{thm:fav}
	The family $\{M_{\alpha}': \alpha=(\theta,1-\theta), \theta\in[0,1]\}$ of G-convex combinations of the Gromov matrices $M_1$ and $M_2$ traces a piecewise linear path from $M_1$ to $M_2$. In particular, it is continuous in $\theta$ in the space of Gromov matrices. 
\end{Theorem}
\begin{IEEEproof}
	See Appendix~\ref{proof:thm:fav}.
\end{IEEEproof}

In the following, we illustrate Theorem~\ref{thm:fav} using a toy example.

\begin{figure}[!htb] 
	\centering
	\includegraphics[width=0.7\columnwidth]{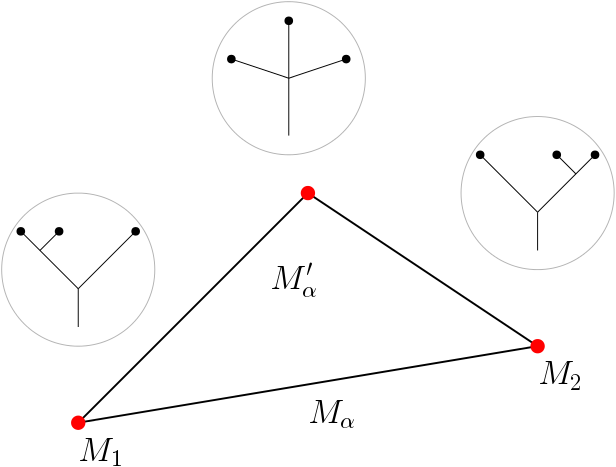}
	\caption{The convex combination $M_{\alpha}$ gives a linear path between the matrices $M_1$ and $M_2$ in the vector space of $n\times n$ matrices. We illustrate the simple case where both $M_1$ and $M_2$ are $3\times 3$ matrices in Example~\ref{eg:ctg}. We also indicate the shape of the corresponding base trees at the turning point and the end points of the piecewise linear path.} \label{fig:9}
\end{figure}

\begin{Example} \label{eg:ctg}
	Consider two Gromov matrices 
	\begin{align*}
	M_1 = \begin{bmatrix}
	d_1 & x_1 & x_2 \\
	x_1 & d_2 &	x_2  \\
	x_2 & x_2 & d_3 \\
	\end{bmatrix}
	\text{ and }
	M_2 = \begin{bmatrix}
	d_1' & x_1' & x_1' \\
	x_1' & d_2' & x_2'  \\
	x_1' & x_2' & d_3' \\
	\end{bmatrix}
	\end{align*}
	such that $x_1 > x_2$ and $x_1' < x_2'$. According to Definition~\ref{def:type}, this is a case when their base sets are of different types (the other case is when the inequalities are reversed and the discussions are similar). For $\theta \in [0,1]$ and $\alpha=(\theta,1-\theta)$, the convex combination of $M_1$ and $M_2$ is given by  
	\begin{align*}
	& M_{\alpha} = \\ 
	& \begin{bmatrix}
	\theta d_1 + (1-\theta) d_1' & \theta x_1+ (1-\theta) x_1' & \theta x_2+(1-\theta) x_1' \\
	\theta x_1+ (1-\theta) x_1'  & \theta d_2 + (1-\theta) d_2'  &	\theta x_2 + (1-\theta) x_2'  \\
	\theta x_2+(1-\theta) x_1' & \theta x_2 + (1-\theta) x_2'  & \theta d_3 + (1-\theta) d_3'  \\
	\end{bmatrix}.
	\end{align*}
	It is clear that $M_{\alpha}(1,3)< \min\{M_{\alpha}(1,2),M_{\alpha}(2,3)\}$. The matrix $M_{\alpha}$ is not a Gromov matrix. To construct the G-convex combination $M_{\alpha}'$, we replace $M_{\alpha}(1,3)$ by $\min\{M_{\alpha}(1,2),M_{\alpha}(2,3)\}.$ It can be shown that $M_{\alpha}(1,2)= M_{\alpha}(2,3)$ if and only if 
	\begin{align*}
	\theta = \theta_* := \frac{x_2'-x_1'}{(x_1-x_2)+(x_2'-x_1')} \in (0,1).
	\end{align*}
	Therefore, for $\theta \in [0, \theta_*]$, we have 
	\begin{align*}
	& M_{\alpha}'= \\
	& \begin{bmatrix}
	\theta d_1 + (1-\theta) d_1' & \theta x_1+ (1-\theta) x_1' & \theta x_2 + (1-\theta) x_2' \\
	\theta x_1+ (1-\theta) x_1'  & \theta d_2 + (1-\theta) d_2'  &	\theta x_2 + (1-\theta) x_2'  \\
	\theta x_2 + (1-\theta) x_2' & \theta x_2 + (1-\theta) x_2'  & \theta d_3 + (1-\theta) d_3'  \\
	\end{bmatrix}
	\end{align*}
	and for $\theta \in [\theta_*, 1]$, 
	\begin{align*}
	& M_{\alpha}'= \\
	& \begin{bmatrix}
	\theta d_1 + (1-\theta) d_1' & \theta x_1+ (1-\theta) x_1' & \theta x_1+ (1-\theta) x_1'  \\
	\theta x_1+ (1-\theta) x_1'  & \theta d_2 + (1-\theta) d_2'  &	\theta x_2 + (1-\theta) x_2'  \\
	\theta x_1+ (1-\theta) x_1'  & \theta x_2 + (1-\theta) x_2'  & \theta d_3 + (1-\theta) d_3'  \\
	\end{bmatrix}.
	\end{align*}
	We see that as $\theta$ increases from $0$ to $1$, $M_{\alpha}'$ moves along a linear path first from $M_1$ to $M_{\theta_*}'$, and then from $M_{\theta_*}'$ to $M_2$. From the expressions of $M_{\alpha}$ and $M_{\alpha}'$, we see that they agree with each other only when $\theta = 0,1$. Therefore, the convex combinations of $M_1$ and $M_2$ for any $\theta \in (0,1)$ are not Gromov. For a graphical illustration, see \figref{fig:9}(a).
\end{Example}

Since the path $\{M_{\alpha}' : \alpha=(\theta,1-\theta), \theta\in[0,1]\}$ is piecewise linear in $\theta$, we can define the \emph{turning points} as those $M_{\alpha}'$ such that in a small neighborhood of which, the path fails to be linear. Inspired by the above example, we make the following observation. 

\begin{Corollary}
	Consider the family $\{M_{\alpha}': \alpha=(\theta,1-\theta), \theta\in[0,1]\}$ of G-convex combinations of the Gromov matrices $M_1$ and $M_2$ and suppose $N_1$ and $N_2$ are two successive turning points with corresponding base trees $T_1$ and $T_2$. Then any triplet $U=\{u_1,u_2,u_3\}$ of distinct vertices in the base set has the same type in $T_1$ and $T_2$.
\end{Corollary}
\begin{IEEEproof}
	Suppose there exists a triplet $U=\{u_1,u_2,u_3\}$ that has different types in $T_1$ and $T_2$. Then as discussed in Example \ref{eg:ctg}, convex combinations of $T_1$ and $T_2$ are not Gromov matrices. Therefore, there must be other turning points between $N_1$ and $N_2$, and it contradicts our assumption that $N_1$ and $N_2$ are successive turning points.
\end{IEEEproof}

The corollary suggests the following intuition: at each turning point, there is a transition of some types of triples, and it is the place where several branches merge together exactly. 

To end this subsection, we make a comparison between the convex combination $M_{\alpha}$ and G-convex combination $M_{\alpha}'$, and discuss the advantages of each approach.

\begin{figure}[!t] 
	\centering
	\includegraphics[width=0.75\columnwidth]{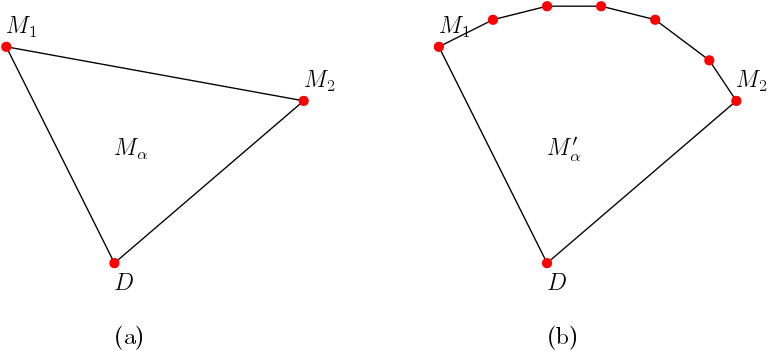}
	\caption{The convex combination ((a)) and G-convex combination ((b)) of two Gromov matrices $M_1,M_2$ and a diagonal Gromov matrix $D$, placed in the vector space of square matrices. In (a), the shape is a genuine planar triangle, while in (b), the shape is not necessary planar. It is because the turning points of the path connecting $M_1$ and $M_2$ may not stay in a common plane.} \label{fig:13}
\end{figure}

First, suppose we consider two general Gromov matrices $M_1,M_2$ and a diagonal one $D$. We form both the convex combination $M_{\alpha}$ and G-convex combination $M_{\alpha}'$. According to Theorem \ref{thm:fav}, if we place them in the space of square matrices, they look like the bounded regions in \figref{fig:13}. The shapes differ slightly, but are still quite similar. If we want to use either $M_{\alpha}$ or $M_{\alpha}'$ to approximate another \emph{matrix} of the same size, the result of the two approaches should be comparable.

Both convex and G-convex combinations have the following advantages in network inference and estimations:
\begin{enumerate}[(a)]
	\item Interpolation with uniform selections of $\alpha$ allows us to select candidates uniformly in the space of symmetric matrices.  
	\item Even if the network topology is missing, estimations can still be performed if a few selected spanning trees are given.
\end{enumerate}
Additional details are provided in Section~\ref{sec:app} in conjunction with applications. Yet, due to differences in the constructions, each approach has its own advantages. The following trade-offs should be taken into consideration:
\begin{enumerate}[(a)]
	\item The computational cost of forming $M_\alpha$ is lower than that for computing $M_\alpha'$.
	\item If the cost function varies smoothly w.r.t.\ the matrix entries, continuous optimization techniques (e.g., convex optimization) can be used to optimize over $M_\alpha$. This is more difficult for $M_\alpha'$ as it does not have an analytical form.
	\item Spectral properties of $M_\alpha$ are better understood, by tools such as the Weyl inequality.  
	\item The main disadvantage of the convex combination $M_\alpha$ is that part of the geometric information is lost; and we are not able to recover any tree in general. On the other hand, for each $\alpha$, the G-convex combination $M_{\alpha}'$ gives a genuine Gromov matrix of a weighted tree. Therefore, the geometric content is preserved and we are able reconstruct the associated tree with ease (see Appendix \ref{geometric} for a discussion.) The reconstruction process has complexity $O(n^3)$, where $n$ is the size of its base set. We point out here that $M_{\alpha}'$ itself may not be a subtree of $G$. However, $M_{\alpha}'$ serves as a good approximation of a spanning tree.
\end{enumerate}

\subsection{A Case Study on Information Propagation Paths}\label{subsec:study}

We analyze and discuss a case study based on information propagation paths to illustrate how well G-convex combinations can approximate a tree of interest. For a graph $G$ and a source node $v\in G$, we generate propagation times on each edge in $G$ independently using a truncated Gaussian distribution. Let $T$ be a shortest path tree from $v$, which can be interpreted as the \emph{information propagation path} from $v$. Let $M$ be the associated Gromov matrix. 

We find a random BFS tree $T_0$ based at $v$. Let $M_0$ be its Gromov matrix and $d_0 = \norm{M-M_0}_2$ be the $L^2$ norm of the difference between $M$ and $M_0$. We choose another BFS tree $T_1$ based at $v$ with the associated matrix $M_1$ and let $D$ be the diagonal matrix of $M_1$. Now we form G-convex combinations between $M_0$ and $M_1$ with parameter $\alpha$ to obtain $M_{\alpha}$; and then form G-convex combinations between $M_{\alpha}$ with $D$ with parameter $\beta$ to obtain $M_{\alpha,\beta}$. Let $d = \min_{\alpha,\beta} \norm{M - M_{\alpha,\beta}}_2 $. We have $d_0\geq d$ and we want to know how large is the ratio $(d_0-d)/d$. A larger ratio means the Gromov method gives a better approximation of $T$.

For the choice of $\alpha$ and $\beta$, we let their components be integer multiples of $1/10$. 
As a consequence, we do not have to deal with too many G-convex combinations. On the other hand, the scale is fine enough such that the convex combinations (being uniformly distributed in the space of symmetric matrices) give reasonably good approximations to an actual spanning tree.

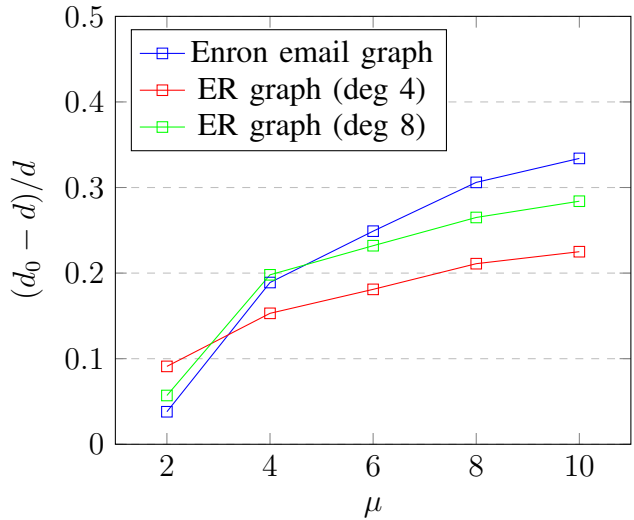
\begin{figure}[!htb]
	\centering
	\begin{tikzpicture}[scale=1]
	\begin{axis}[
	xlabel={$\mu$},
	ylabel={$(d_0-d)/d$},
	xmin=1, xmax=11,
	ymin=0, ymax=0.5,
	xtick={2,4,6,8,10},
	ytick={0,0.1,0.2,0.3,0.4,0.5},
	legend pos=north west,
	ymajorgrids=true,
	grid style=dashed,
	]
	
	\addplot[
	color=blue,
	mark=square,
	]
	coordinates {
		(2,0.038)(4,0.189)(6,0.249)(8, 0.306)(10, 0.334)
	};
	
	\addplot[
	color=red,
	mark=square,
	]
	coordinates {
		(2,0.091)(4,0.153)(6,0.181)(8, 0.211)(10, 0.225)
	};
	
	\addplot[
	color=green,
	mark=square,
	]
	coordinates {
		(2,0.057)(4,0.198)(6,0.232)(8, 0.265)(10, 0.284)
	};
	
	\legend{Enron email graph, ER graph (deg 4), ER graph (deg 8)}
	
	\end{axis}
	\end{tikzpicture}
	\caption{Plot of the ratio $(d_0-d)/d$ against propagation mean $\mu$.} \label{fig:11}
\end{figure}

\begin{figure}[!htb] 
	\centering
	\includegraphics[width=0.5\textwidth]{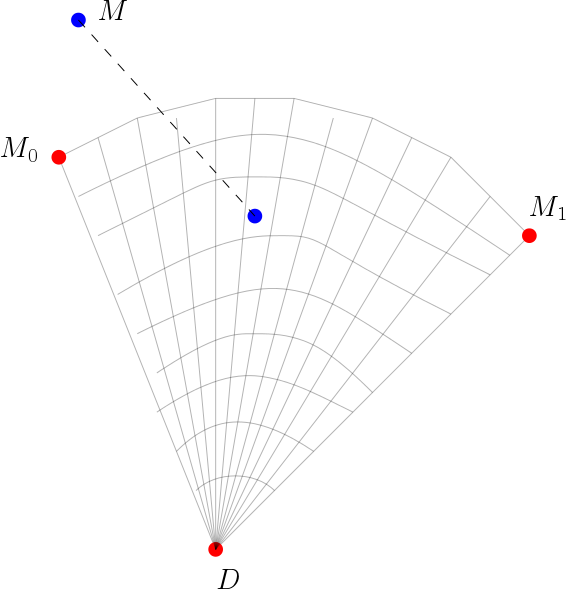}
	\caption{Illustration of the explanation of the observation regarding \figref{fig:11}.} \label{fig:12}
\end{figure}

We perform simulations on both synthetic Erd\"{o}s-R\'{e}nyi (ER) graphs and the Enron email network\footnote{https://snap.stanford.edu/data/email-Enron.html} (as an example of a scale-free graph) with $500$ nodes. For convenience, we fix the variance $\sigma^2$ of the truncated Gaussian distribution as $1$ and vary the mean $\mu$. Intuitively, if the mean is larger, the information propagation path $T$ looks more like a BFS tree. The simulation results are plotted in \figref{fig:11}. We observe that if $\mu$ is larger (i.e., the actual information propagation path $T$ resembles a BFS tree more closely), then our approach gives a much better approximation of $T$ than simply choosing a random BFS tree $T_0$. The intuition is as follows: even though the randomly chosen BFS trees $T_0,T_1$ are not close to $T$, there exists a projection of $M$ in the convex region formed by $M_0$, $M_1$, and $D$ (see \figref{fig:12}). This example suggests that by using G-convex combinations to approximate the information propagation path $T$ in problems where $T$ is unknown \emph{a priori} is a better approach than the spanning tree heuristic.

\section{Applications of the Gromov Method} \label{sec:app}

The theory developed in this paper can be used as an alternative to the spanning tree heuristic in network inference and estimation problems proposed by various works like \cite{Shah2011, Pinto2012, Zhu2013, LuoTayLen14, JiTayVar:J17, Luo18}. Suppose we are given a graph $G=(V,E)$, a subset of nodes $U\subset V$, and an estimator or cost function $\phi(T,s)$ taking as inputs a tree $T$ spanning $U$ and base vertex $s$, with $(T,s)$ satisfying certain criteria according to the problem of interest. We denote by $\mathcal{S}_0$ the set of all $(T,s)$ satisfying these criteria. An inference problem typically involves computing  
\begin{align}\label{inf_exp}
\mathbb{E}_{(T,s)\in\mathcal{S}_0}[\phi(T,s)],
\end{align}
where the expectation is w.r.t. a distribution over $(T,s)\in\mathcal{S}_0$, or 
\begin{align}\label{inf_opt}
\min_{(T,s)\in \mathcal{S}_0}\phi(T,s).
\end{align}
As the cardinality of $\mathcal{S}_0$ is often very large, the spanning tree heuristic approximates the solution by randomly choosing a small subset $\mathcal{S}_1\subset\mathcal{S}_0$ and computing either $\mathbb{E}_{(T,s)\in\mathcal{S}_1}[\phi(T,s)]$ or $\min_{(T,s,U)\in \mathcal{S}_1}\phi(T,s)$ instead. For our \emph{Gromov method}, instead of randomly selecting a spanning tree, we let $\mathcal{S}_2$ be obtained by taking either convex or G-convex combinations of the trees in $\mathcal{S}_1$. For example, for every $T_1$ and $T_2$ with associated Gromov matrices $M_1$ and $M_2$ in $\mathcal{S}_1$, we can let $D$ be the diagonal matrix taking the diagonal of $M_1$ and form ${\bf M}=\{M_1,M_2,D\}$. We then form either a convex combination or G-convex combination from ${\bf M}$. All such combinations are then included in $\mathcal{S}_2$. We then compute $\mathbb{E}_{(T,s)\in\mathcal{S}_2}[\phi(T,s)]$ or $\min_{(T,s,U)\in \mathcal{S}_2}\phi(T,s)$. It is clear that the Gromov method achieves a better approximation since $\mathcal{S}_1\subset\mathcal{S}_2$, but requires higher computational complexity. We remark here that as a general suggestion, when we form G-convex combination, it usually suffices to choose each weight parameter as integer multiples of $1/10$ or $1/20$ in $[0,1]$.

The advantages of the Gromov method are as follows:
\begin{enumerate}[(a)]
	\item Taking the convex or G-convex combination gives us a larger family of approximate spanning trees, while retaining certain structural features of the ambient graph $G$ (see Corollary~\ref{coro:lmam}, \figref{fig:4} and Proposition~\ref{prop:ssia}). Standard optimization techniques may then be applied to optimize $\phi$ over this family. For the convex combination approach, we may use continuous optimization tools. For $G$-convex combinations, we perform a discrete optimization with fixed size increments on the combination weights. 
	\item In some applications, we may not have access to all the spanning trees satisfying the problem criteria, or it may be computationally complex or practically expensive to find such spanning trees. It becomes difficult to construct many spanning trees using the tree heuristic. With the Gromov method, we mitigate this problem by using only a small set of spanning trees and then forming convex or G-convex combinations from them.
\end{enumerate}

In the following, we present several network inference applications to illustrate the use of the Gromov method.

\subsection{Information Acquisition Order}\label{sec:inf}

Suppose that information is propagated stochastically from a source node $s$ in a graph $G$, where the propagation time distribution along each edge is known. An important application is knowing if on average a node $u$ lies on the propagation path from $s$ to another node $v$. For example, in viral marketing and social learning (e.g., \cite{Dom2005, Leskovec2007, SohTayQue:J13, Tay:J15}), we are interested to find those nodes in a social network that act as the gatekeepers of information propagation. In infection control and monitoring (e.g., \cite{LuoTayLen:J16, pas01, Dez01, pas02, wang09}), we are interested to identify those nodes in a network with the highest probability of acquiring the infection and spreading it to others in the network so that we can add monitoring and countermeasures at these nodes.

Information propagation from $s$ follows a random spanning tree $T$. Computing the probability that $u\in [s,v]$, where $[s,v]$ is the unique path in $T$ connecting $s$ and $v$ is intractable for a dense graph $G$. Furthermore, such a computation requires full knowledge of the propagation statistics and graph topology. To estimate this probability empirically, we can sample a large number of random spanning trees containing $s, u$ and $v$. For each such spanning tree $T$, we decide if $u$ is on the unique path $[s,v]$ connecting $s$ and $v$ or not. We can then compute the \emph{empirical probability} of instances of $u \in [s,v]$. This problem is of the form Problem~(\ref{inf_exp}), where $\phi(T,s)$ is the indicator function of the event that $u\in[s,v]$.

In actual applications, we may not have the resources to sample so many random spanning trees. For example, in viral marketing, each sample involves an information dissemination campaign, which can be expensive to repeat. In the Gromov method, we may sample a very small number of spanning trees, and use G-convex combinations to synthesize more spanning trees. 

It is interesting to note that to perform this estimation, we do not even need to know the network topology, provided we are given a few samples. This is a very useful feature for practical applications as inferring the full network topology is usually not an easy task. In this problem, we need to use G-convex combinations, as geometric information of the synthesized trees is used. 

We performed simulations on four types of graphs: 2D-lattice, the Enron email network, a sample Facebook network\footnote{https://snap.stanford.edu/data/egonets-Facebook.html} and the complete graph, as an extreme case. For each simulation trial, a random sample is generated as follows: (1) propagation timestamps across the network are simulated; (2) a minimum spanning tree is then chosen by Dijkstra's algorithm. For the Gromov method, we sample three spanning trees and form $10$ G-convex combinations between each pair of sampled spanning trees by choosing $\theta$ in the combination weight vector $\alpha=(\theta,1-\theta)$ uniformly from $[0,1]$ in integer multiples of $1/10$, giving approximately $30$ synthesized trees. The ground truth is obtained by averaging over $200-500$ samples depending on the density of the network.

The accuracy of the Gromov method is then computed as follows: For each triple of nodes $\{s,u,v\}$, from the ground truth, we may find the empirical probability of (1) $u\in [s,v]$, (2) $u\notin [s,v]$ as $p_1$ and $p_2=1-p_1$. On the other hand, the Gromov method gives estimations $\hat{p}_1$ and $\hat{p}_2$. The error $e_{\{s,u,v\}}$ is calculated as $(|p_1-\hat{p}_1|/p_1+|p_2-\hat{p}_2|/p_2)/2$. The accuracy is obtained by averaging $1-e_{\{s,u,v\}}$ over different triples $\{s,u,v\}$.

It is useful to notice that the size of the network is not a crucial factor here since we only need to make sure the spanning tree is large enough to contain the three involved nodes. On the other hand, the ratio $r$ between the average node degree and the number of nodes gives an important indication of the density of the graph. The simulation results are shown in Table~\ref{tab:1}. We see that the Gromov method performs reasonably well and yields a good estimation with a limited number of samples.
\begin{table}[!htbp]
	\centering  
	\begin{tabular}{|l|c|c|}  
		\hline
		&Degree-node ratio $r$ &Accuracy\\ 
		\hline  
		2D-lattice & $<$0.04 & 95.1\% \\
		\hline
		Email network & 0.094 & 95.3\%   \\
		\hline        
		Facebook network & 0.268 & 91.1\%  \\   
		\hline
		Complete graph & $\approx$ 1 & 94.0\% \\
		\hline
	\end{tabular}
	\caption{Accuracy of the Gromov method in estimating information acquisition order.}
	\label{tab:1}
\end{table}

\subsection{Network Source Identification}\label{subsec:source}

We consider the problem posed in Example~\ref{example:infection} in the introduction. We start with the case where a snapshot of the infection status of each node is observed. We assume that infection propagation along each edge follows an exponential distribution with mean $1$. Suppose we are given a snapshot $U\subset V$ of all the infected nodes. We want to estimate the identity of the source node. One possible approach is estimator based. An estimator $e(s,T_s)$ is a real valued function on each candidate source node $s$ and a spanning tree $T_s$ of $I$ rooted at $s$. The source is found by maximizing $e(\cdot,\cdot)$ over all $s$. In the spanning tree heuristic adopted by \cite{Shah2011}, for a general graph, a random BFS tree $T_s$ is usually chosen as the second argument of $e(\cdot,\cdot)$, which might be a source of error in the estimation. 

Various estimators have been studied in the literature \cite{Shah2011, LuoTayLen14b, Tan2016, ZhuY2016}, and for demonstration purposes, we choose the centroid $e_C(\cdot,\cdot)$ as an example. The centroid $e_C(s,T)$ (see \cite{Tan2016}) is defined as the maximal size, i.e., the sum of all the edge weights, of connected components of $T\backslash \{s\}$. This problem is of the form Problem~(\ref{inf_opt}), where we seek to minimize the cost function $\phi(T,s)=-e_C(s,T)$ over all trees $T$ rooted at $s$ that span the observed infected nodes $U$. In the spanning tree heuristic, this optimization is done in the following way:
\begin{enumerate}[(a)]
	\item For each candidate source $s\in V$, we randomly choose a BFS tree $T_{\mathrm{BFS}}(s)$. 
	\item The infection source is then estimated by $\displaystyle\text{argmax}_{s\in V} e_C(s,T_{\mathrm{BFS}}(s))$.
\end{enumerate}
In the Gromov method, for each base vertex $s$, we first fix an ordering of the nodes. We choose the BFS tree $T_1$ (with $M_1$ its Gromov matrix) according to this ordering and the BFS tree $T_2$ with the reverse node ordering. Let $T_3$ be the tree associated with the diagonal matrix of $M_1$, for edge weights scaling. We use $\{T_1,T_2,T_3\}$ as the initial spanning trees to form G-convex combinations $\mathcal{S}_2(s)$ by setting the combination weight vector $\alpha=(\alpha_1,\alpha_2,\alpha_3)$ so that each $\alpha_i$ is an integer multiple of $1/10$ chosen uniformly in $[0,1]$ with $\sum_i \alpha_i = 1$. We then estimate the source by $\displaystyle\text{argmax}_{s\in V, T\in\mathcal{S}_2(s)} e_C(s,T)$.

\begin{table}[!htb]
	\centering  
	\begin{tabular}{|l|c|c|c|c|}  
		\hline
		&Ave. Deg. & Size &Err. Red. & Detect. Improve.\\ 
		\hline  
		BA  & 4 & 500 & 12.4\% & 73.7\% \\
		\hline
		ER  & 4 & 500 & 18.6\% & 25.8\% \\
		\hline
		Email  & 9.86 & 670 & 12.7\% & 25.0\%  \\
		\hline    
		Facebook  & 35.7 & 786 & 2.1\% & 29.4\% \\   
		\hline 
	\end{tabular}
	\caption{Error reduction and detection improvement of the Gromov method over the spanning tree heuristic in network source estimation.} \label{tab:2}
\end{table}    

We run simulations on both synthetic and real networks: Barab\'{a}si-Albert (BA) graphs with $500$ vertices and degree $4$, ER graphs with $500$ vertices and average degree $4$, the Enron email network with $670$ vertices and average degree $9.86$, and the Facebook network with $786$ vertices and average degree $35.7$. In each simulation, we infect $20\%$ to $30\%$ of the vertices where the infection propagation time from an infected node to an uninfected neighbor follows an exponential distribution with rate 1. Let the average distance between the true source and the estimated source be $d_{B}$ for the BFS heuristic and $d_G$ for the Gromov method. The error reduction rate shown in Table~\ref{tab:2} is given by $(d_{B}-d_G)/d_B$. We also compute the fraction of trials for which the source is among the top $20\%$ highest scoring nodes determined by each method (i.e., the \emph{$20\%$-accuracy}). Let this be $p_B$ for the BFS heuristic and $p_G$ for the Gromov method. We compute the \emph{detection improvement} as $(p_G-p_B)/p_B$ in Table~\ref{tab:2}. We see noticeable improvement when using the Gromov method in most of the cases. The error reduction for the Facebook network is smaller because the network is very dense. However, we see significant improvement in the $20\%$-accuracy score.

\begin{figure}[!htbp]
	\footnotesize
	\centering
	\begin{minipage}[b]{.5\linewidth}
		\centering
		\centerline{\includegraphics[width=5.5cm]{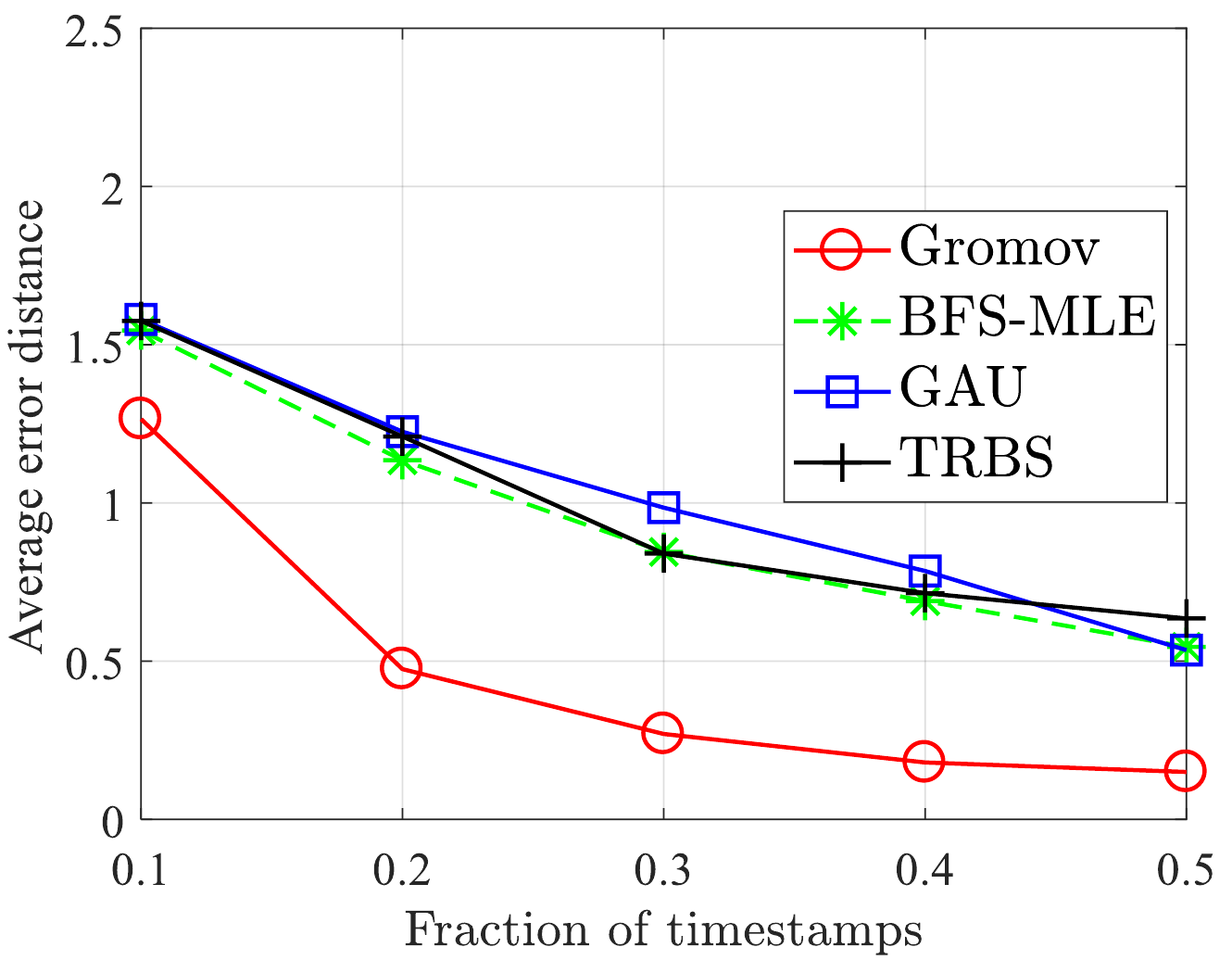}}
		\centerline{(a) BA graph}
	\end{minipage}%
	\begin{minipage}[b]{.5\linewidth}
		\centering
		\centerline{\includegraphics[width=5.5cm]{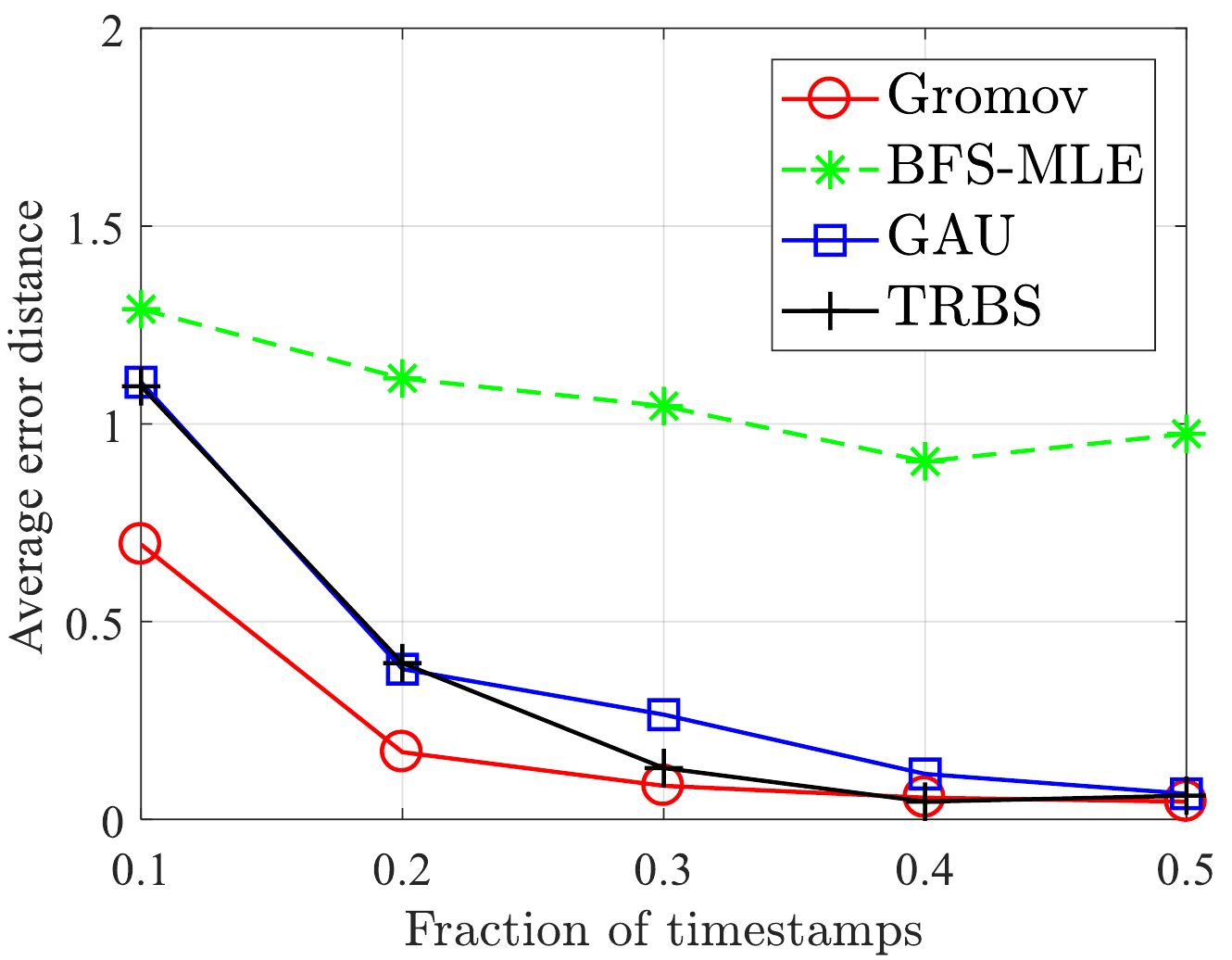}}
		\centerline{(b) ER graph}
	\end{minipage}
	\begin{minipage}[b]{.5\linewidth}
		\centering
		\centerline{\includegraphics[width=5.5cm]{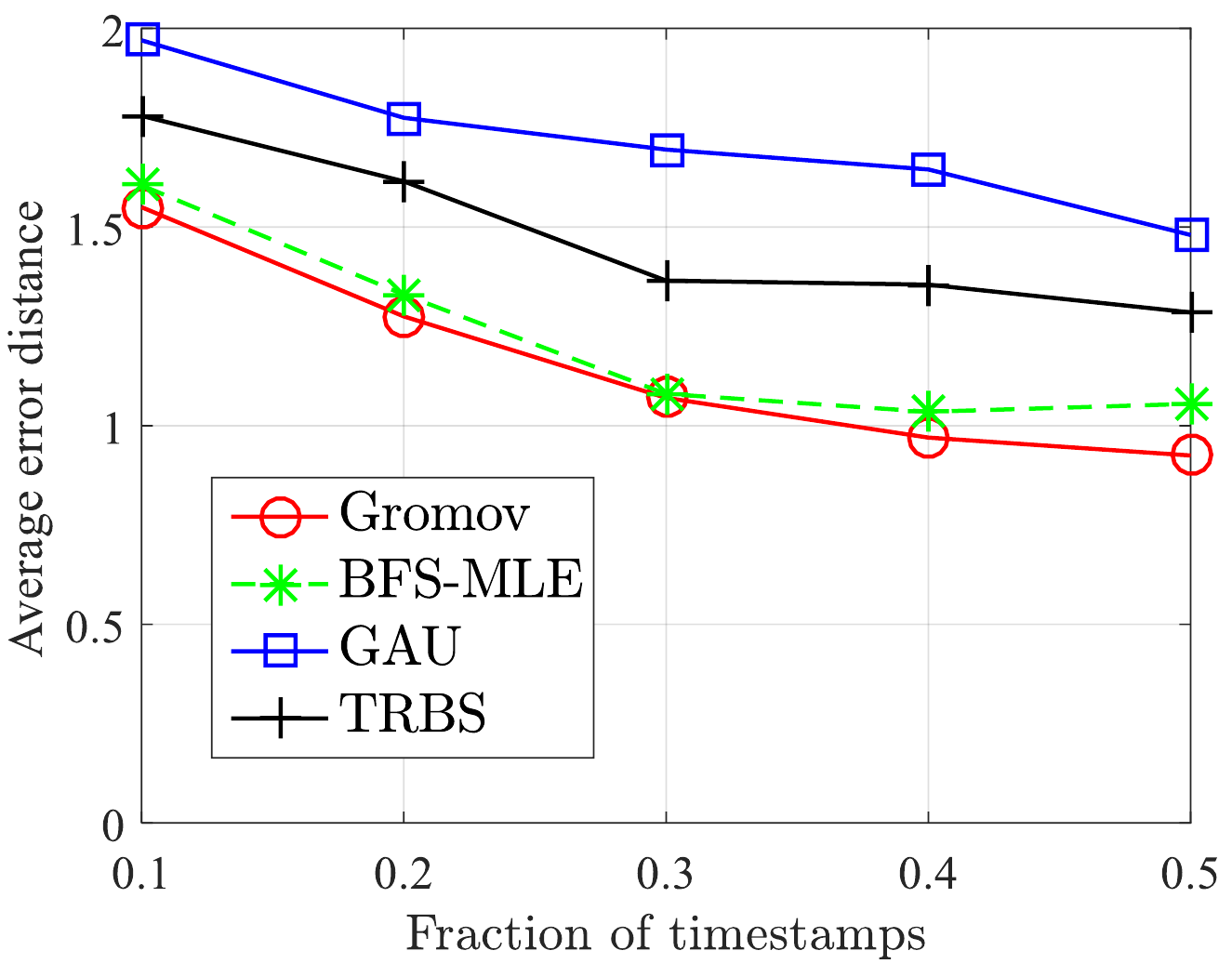}}
		\centerline{(c) Email}
	\end{minipage}%
	\begin{minipage}[b]{.5\linewidth}
		\centering
		\centerline{\includegraphics[width=5.5cm]{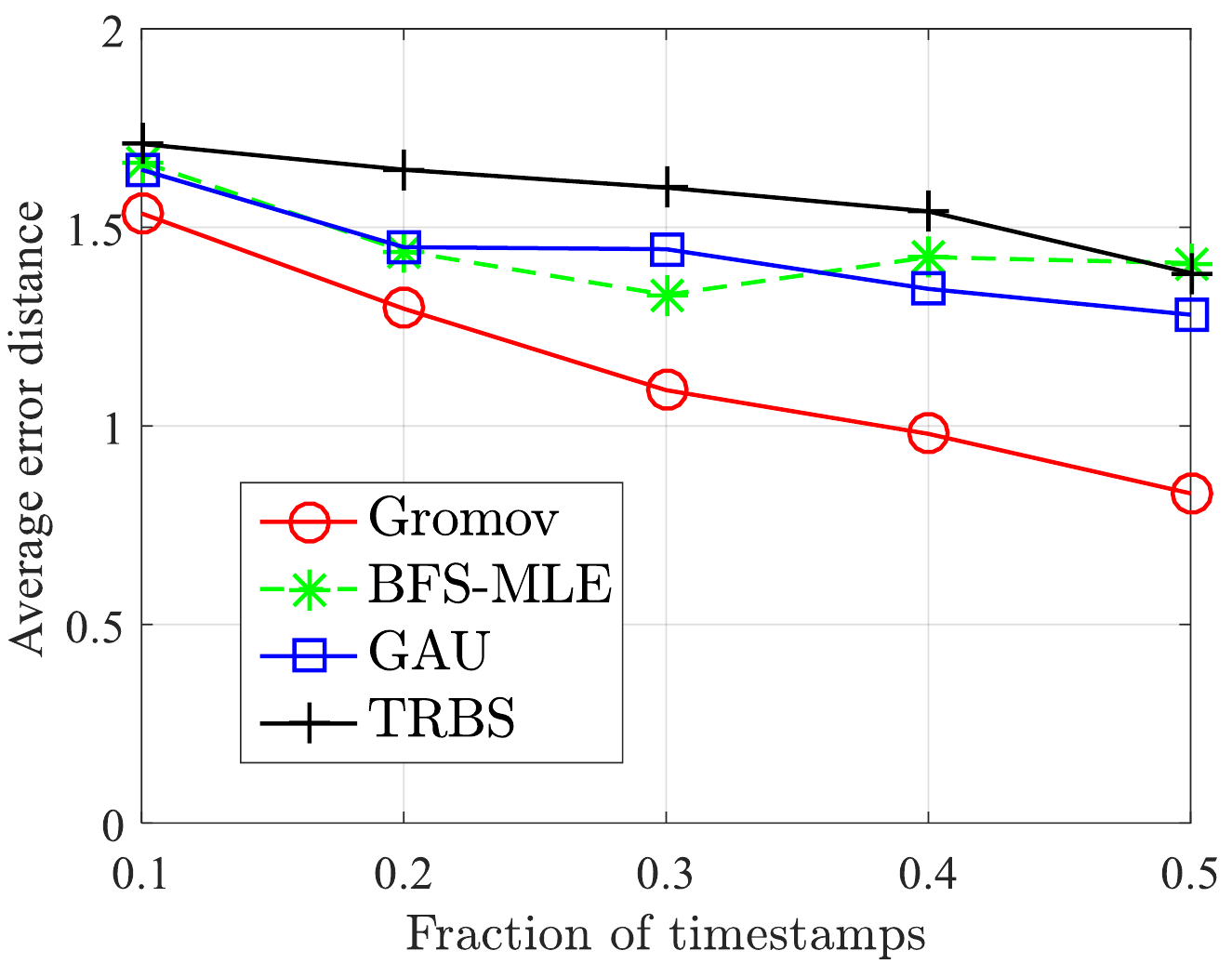}}
		\centerline{(d) Facebook}
	\end{minipage}
	\caption{Performance comparison (average error distance) for the network source estimation problem with timestamps.}
	\label{fig:16}
\end{figure} 

\begin{figure}[!htbp]
	\footnotesize
	\centering
	\begin{minipage}[b]{.5\linewidth}
		\centering
		\centerline{\includegraphics[width=5.5cm]{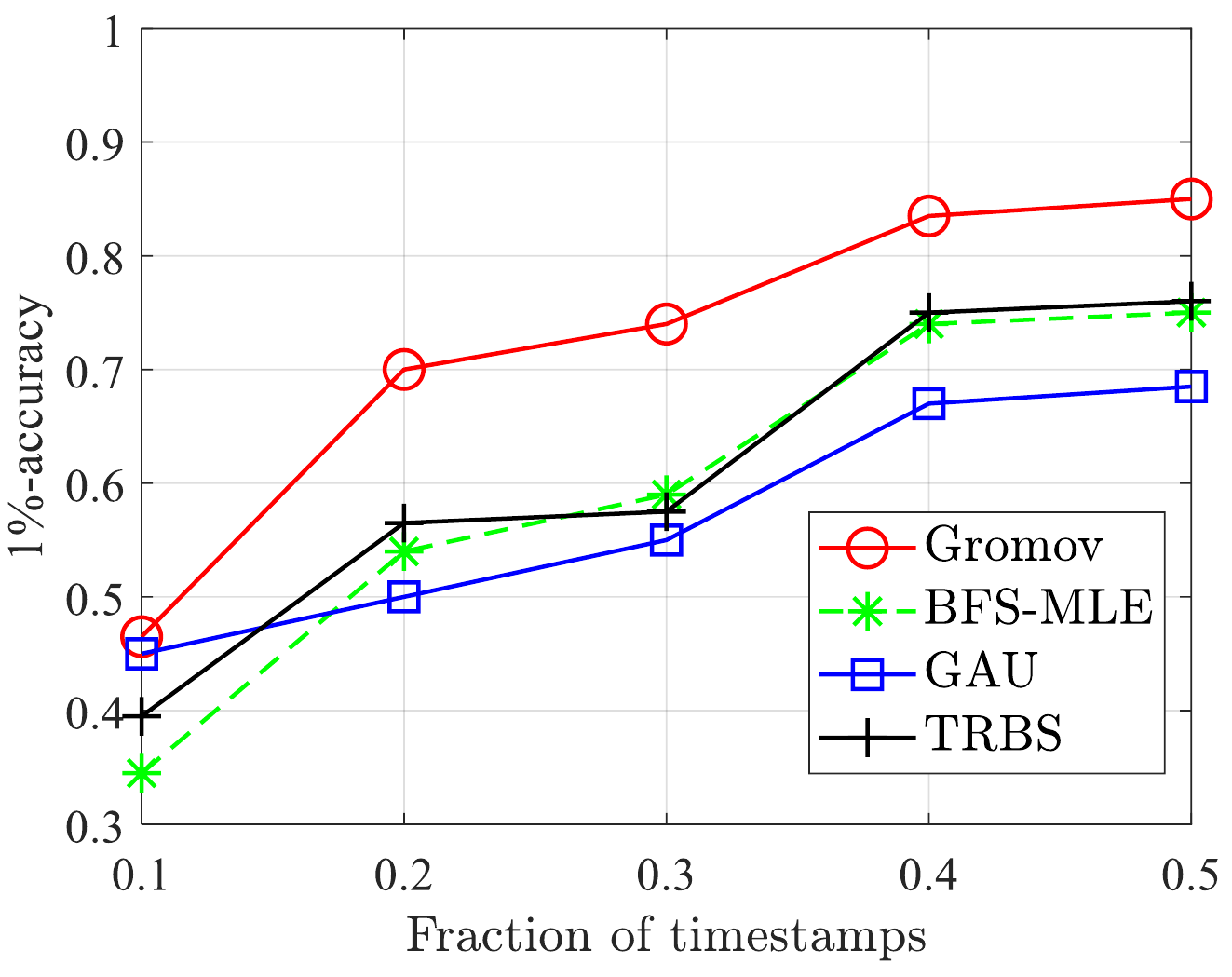}}
		\centerline{(a) BA graph}
	\end{minipage}%
	\begin{minipage}[b]{.5\linewidth}
		\centering
		\centerline{\includegraphics[width=5.5cm]{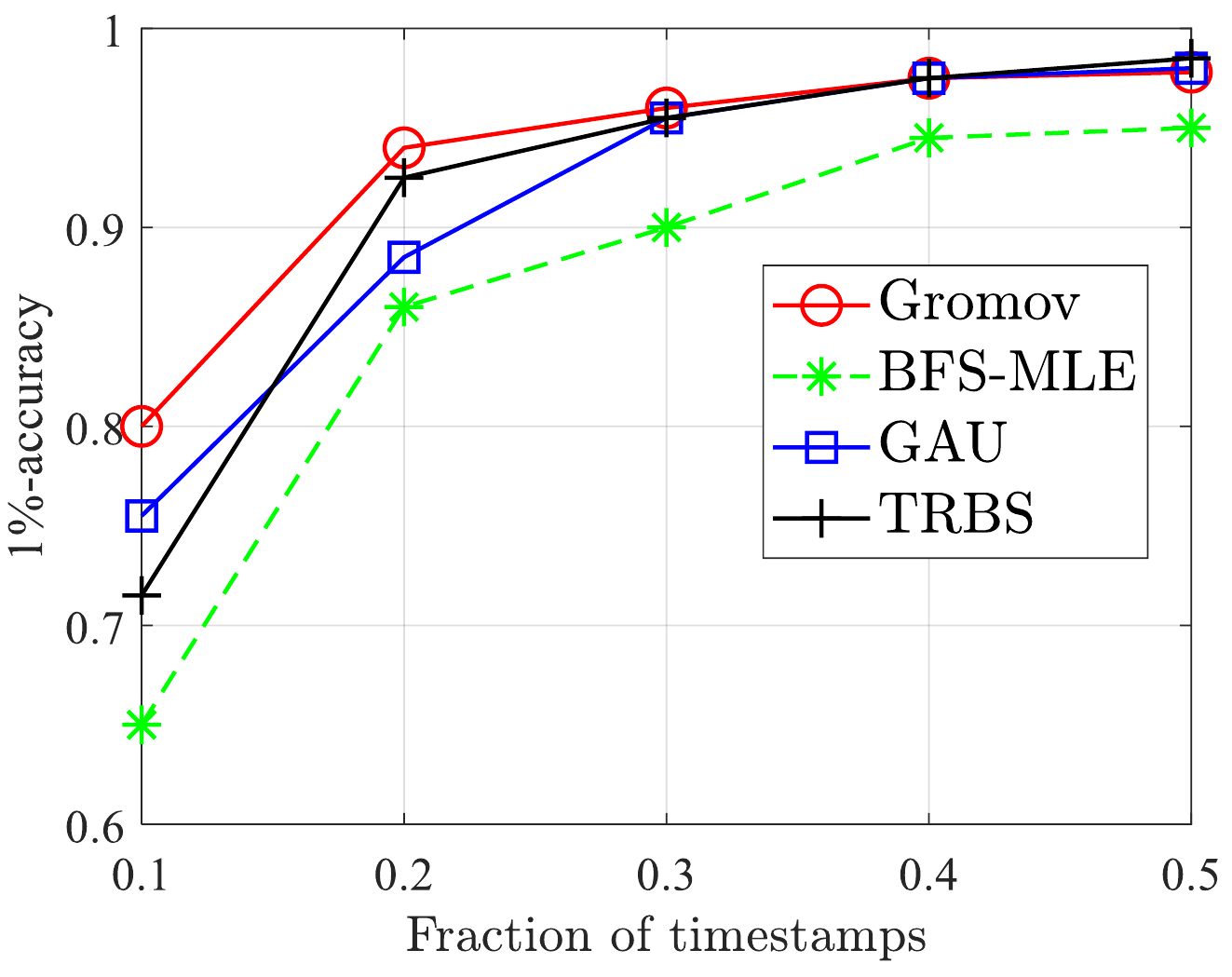}}
		\centerline{(b) ER graph}
	\end{minipage}
	\begin{minipage}[b]{.5\linewidth}
		\centering
		\centerline{\includegraphics[width=5.5cm]{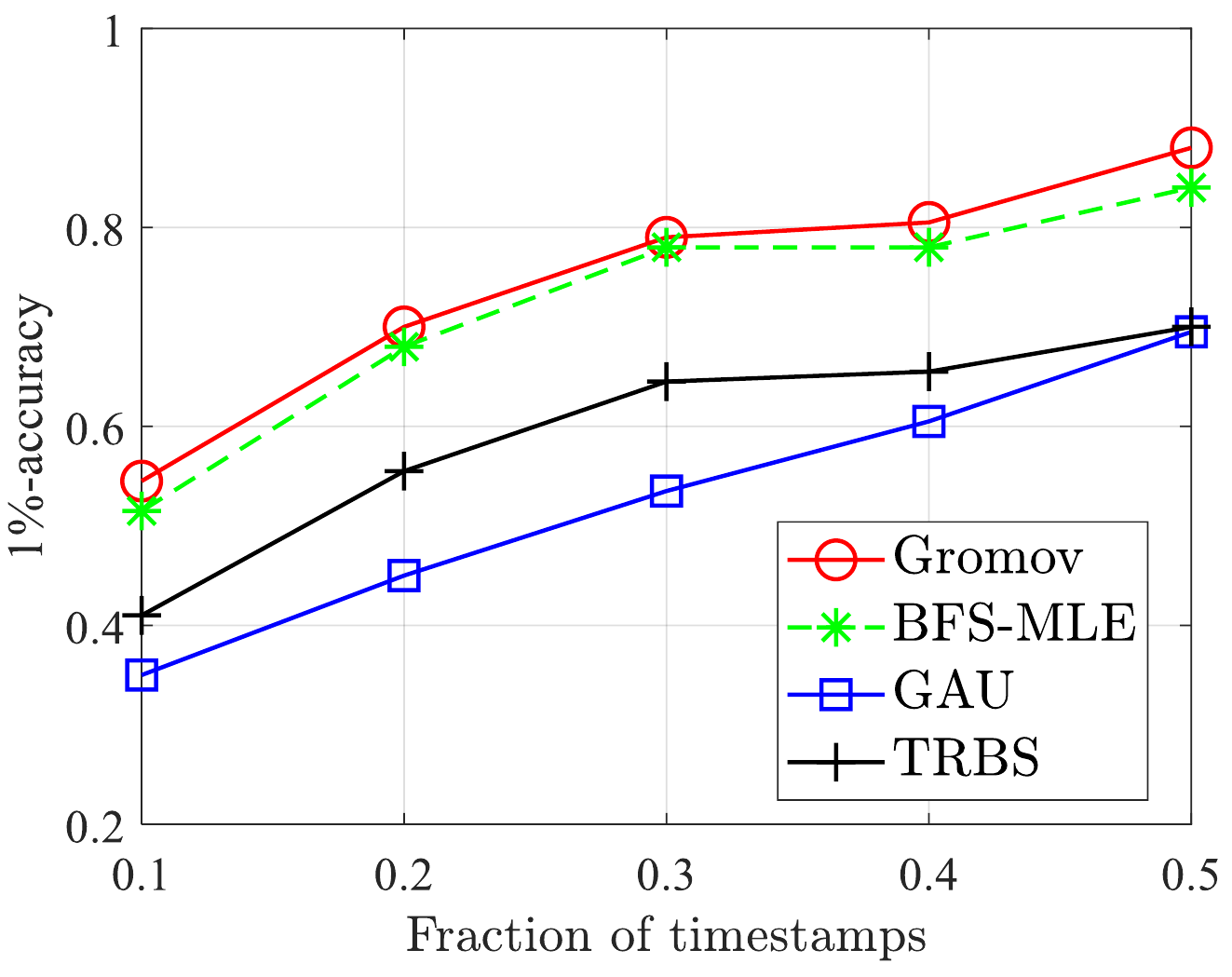}}
		\centerline{(c) Email}
	\end{minipage}%
	\begin{minipage}[b]{.5\linewidth}
		\centering
		\centerline{\includegraphics[width=5.5cm]{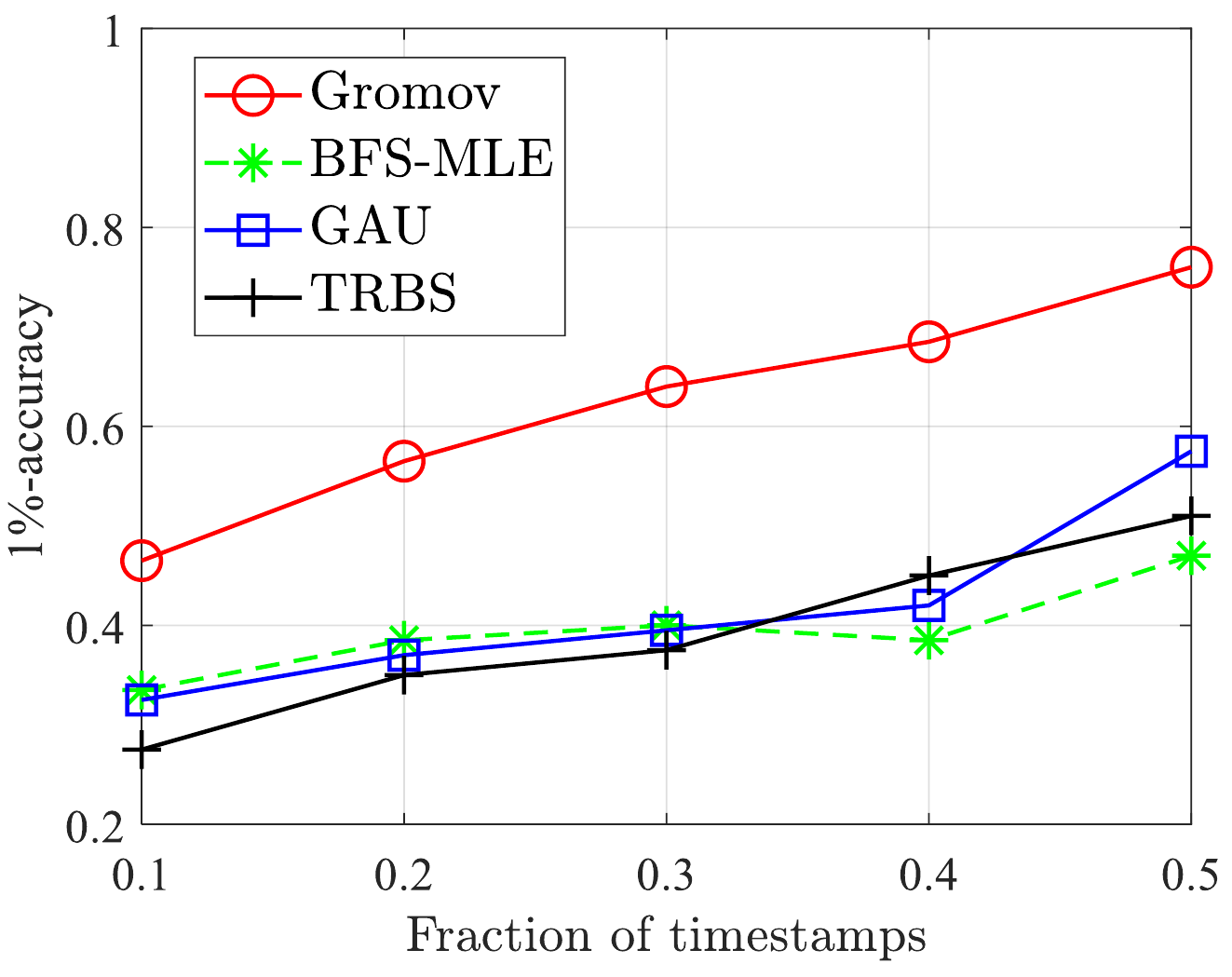}}
		\centerline{(d) Facebook}
	\end{minipage}
	\caption{Performance comparison ($1\%$-accuracy) for the network source estimation problem with timestamps.}
	\label{fig:18}
\end{figure} 

Another variant of the problem is source estimation with observed infection timestamps. Here, we assume that a small percentage of the infected nodes are observed, together with their respective infection timestamps. We apply the Gromov method (using convex combination), similar to the snapshot observation based estimation described above (we refer the reader to \cite{Tan18} for details). We compare the Gromov method with the BFS tree based approach (called BFS-MLE), as well as other approaches: GAU \cite{Pinto2012} and TRBS \cite{Shen2016}, on various networks. We use average error distance as well as $1\%$-accuracy as the evaluation metrics. The results are shown in \figref{fig:16} and \figref{fig:18}, which are reproduced from \cite{Tan18} here for completeness. We vary the number of nodes with observed timestamps. We observe significant performance improvement using the Gromov method in most cases.

\subsection{Data Center Placement}

In this application, we consider the problem of determining the optimal placement of data centers in a large network \cite{Li1999,Jain2001,Charikar2002,Arya2004,Laoutaris2007,Oikonomou2012,Qiu01}. The network is given by a weighted graph $G=(V,E)$. For each node $v \in V$, a non-negative number $w(v)$ is associated to it to indicate the service demand at $v$. Given a subset $S\subset V$ and $v \in V$, let $\rho(v,S) = \min_{s\in S} d_G(s,v),$ i.e., path length from $v$ to the node in $S$ closest to $v$. The overall weighted cost of serving $V$ by $S$ is defined as $\text{cost}(S) = \sum_{v\in V}w(v)\rho(v,S)$. We want to find a subset $\hat{S} = \argmin_{S\subset V, |S|=k}\text{cost}(S)$, where $k$ is the number of data centers. This problem is similar to Problem~(\ref{inf_opt}), except that now we are finding an optimal subset of nodes.

An iterative greedy algorithm was proposed by \cite{Qiu01} to solve this problem as follows: Let $\hat{S}_{0}=\emptyset$. In each iteration $1\leq l \leq k$, suppose that a set of data centers $\hat{S}_{l-1}$ with $l-1$ nodes has been found. Find the node $\hat{s} = \text{argmin}_{s\in V}\text{cost}(\hat{S}_{l-1}\cup\{s\})$ and let $\hat{S}_{l}=\hat{S}_{l-1}\cup\{\hat{s})\}$. Note that the greedy algorithm may not find the optimal solution.

We propose a Gromov method based on the algorithm in \cite{Luo18}, which we enhance using the Gromov techniques discussed in this paper. The steps are as follows: 
\begin{enumerate}[(a)]
	\item Randomly choose an initial set of $k$ nodes $\hat{S}_0=\{s_0^1,\ldots,s_0^k\}$.
	\item For each iteration $l\geq 1$, partition $G$ using a Voronoi partition $G_1\cup\ldots\cup G_k$ with vertices from $\hat{S}_{l-1}=\{s_{l-1}^1,\ldots,s_{l-1}^k\}$ as the centers. For each Voronoi partition or subgraph $G_i$, find a minimum spanning tree and a BFS tree rooted at its center $s_{l-1}^i$. Let $\mathcal{T}_i$ be the set of trees formed by the G-convex combination of these two spanning trees, where we choose the combination weight vector $\alpha=(\theta,1-\theta)$ so that each $\theta$ is an integer multiple of $1/10$ chosen uniformly in $[0,1]$. Solve for $\min_{s\in G_i, T\in\mathcal{T}_i} \sum_{v\in G_i} w(v) d_{T}(s,v)$ (using Algorithm~2 in \cite{Luo18}) to find $s_l^i$. Set $\hat{S}_{l}=\{s_{l}^1,\ldots,s_{l}^k\}$.
	\item Iterate till $\max_{1\leq i \leq k} d_G(s_l^i, s_{l-1}^i) \leq \eta$ for some small $\eta>0$.
\end{enumerate}

Intuitively, the Gromov method attempts to evenly distribute the data centers across the network. Therefore, a improvement over the greedy approach is possible, as the latter may result in some data centers being ineffectively placed.

We run simulations on both grid networks and CAIDA AS graphs (see \cite{Leskovec2005}). We vary both the graph size ($100$ to $400$ nodes) and number of data centers ($1\%, 2\%$ of the total number of nodes) for each graph type. The service demand for each node is generated according to the Pareto distribution, which is a power-law distribution used to model the situation where a small number of nodes generate most of the service demand. 

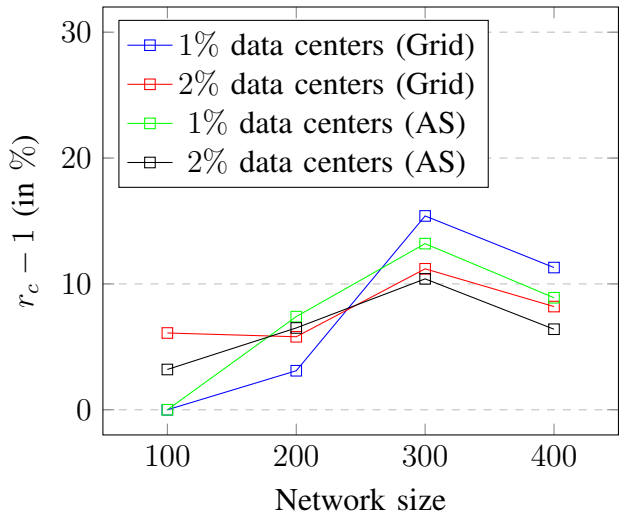
\begin{figure}[!htb]
	\centering
	\begin{tikzpicture}[scale=1]
	\begin{axis}[
	xlabel={Network size},
	ylabel={$r_c-1$ (in $\%$)},
	xmin=50, xmax=450,
	ymin=-2, ymax=32,
	xtick={0,100,200,300,400},
	ytick={0,10,20,30},
	legend pos=north west,
	ymajorgrids=true,
	grid style=dashed,
	]
	
	\addplot[
	color=blue,
	mark=square,
	]
	coordinates {
		(100,0)(200,3.1)(300,15.4)(400,11.3)
	};
	
	\addplot[
	color=red,
	mark=square,
	]
	coordinates {
		(100,6.1)(200,5.8)(300,11.2)(400,8.2)
	};
	
	\addplot[
	color=green,
	mark=square,
	]
	coordinates {
		(100,0)(200,7.4)(300,13.2)(400,8.9)
	};
	
	\addplot[
	color=black,
	mark=square,
	]
	coordinates {
		(100,3.2)(200,6.5)(300,10.4)(400,6.4)
	};
	\legend{$1\%$ data centers (Grid), $2\%$ data centers (Grid), $1\%$ data centers (AS), $2\%$ data centers (AS)}
	
	\end{axis}
	\end{tikzpicture}
	\caption{Performance comparison between the Gromov method and the greedy method on grid networks and AS graphs for the data center placement problem.} \label{fig:10}
\end{figure}

For each instance, we compute the estimated cost $c_1$ for the greedy method, and $c_2$ for the Gromov method. The performance of the two methods are compared by taking the average of the ratio $r_c=c_1/c_2$. The simulation results are shown in \figref{fig:10}. We see that the Gromov method has a better performance in almost all the cases (average cost ratio $r_c-1>0$).    

\section{Conclusion} \label{sec:con}

In this paper, using the Gromov matrix interpretation of trees, we have proposed a modified version of the commonly used spanning tree heuristic, which we call the Gromov method, in network inference problems. We presented both necessary and sufficient conditions for a matrix to be a Gromov matrix of a weighted tree, followed by a demonstration on how to form convex combination of Gromov matrices. The Gromov method proposed is based on optimizing the network inference objective function over a parametrized family of Gromov matrices. We give some applications of the Gromov method on network inference and estimation. As the proposed Gromov method provides a general framework, it is of interest to explore other network inference and estimation applications in future works. Moreover, how to choose the appropriate matrices or spanning trees in forming the convex combinations is also an interesting and important topic to explore.

\appendices

\section{Proof of Proposition~\ref{prop:mit}}\label[appendix]{proof:prop:mit}

Suppose that the base of $M$ is $(T,s,V)$. We first note that conjugation by permutation matrices is equivalent to a re-ordering of the indices of $M$. Therefore, we can ignore the ordering of the indices in our proof. To prove the proposition, it suffices to prove that the weighted tree with base $(T, s,V)$ can be obtained from the Gromovication operations starting from $(\{s\}, s, \emptyset)$. 

\begin{figure}[!tb] 
	\centering
	\includegraphics[width=0.95\columnwidth]{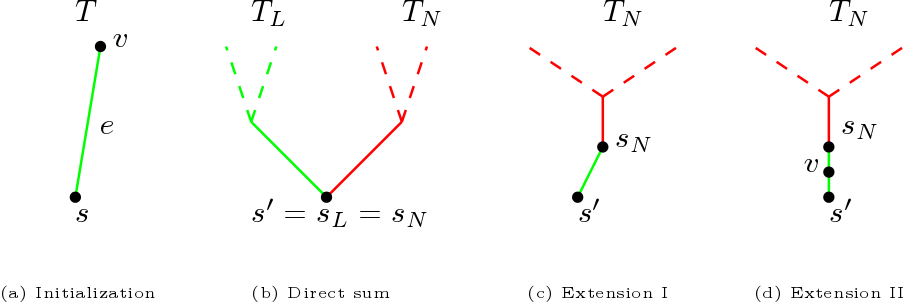}
	\caption{Illustration of the geometric constructions corresponding to the Gromovication operations.} \label{fig:5}
\end{figure}

We first observe that each Gromovication operation on Gromov matrices corresponds to the construction of a new base from the bases associated with the input Gromov matrices as follows (see \figref{fig:5} for an illustration).
\begin{itemize}
	\item Initialization: Given the base $(\{s\},s,\emptyset)$ and $a>0$, we form a tree by attaching an edge $e$ of weight $a$ to $s$. Denote the other end of $e$ by $v$ and let $T = (\{s,v\},e)$. Then the base of the initialization is $(T,s,\{v\})$.
	\item Direct sum: Given the bases $(T_L,s_{L},V_{L})$ and $(T_N,s_{N},V_{N})$ corresponding to the Gromov matrices $L$ and $N$ respectively, we form a new tree $T'$ by joining the two trees $T_L$ and $T_N$ at their base vertices and identifying both $s_L$ and $s_N$ as a single base vertex $s'$. A base of $L\oplus N$ is then $(T', s', V_{L}\cup V_{N})$.		
	\item Extension I (with parameter $a>0$): Given a base $(T_N, s_{N}, V_{N})$ corresponding to the Gromov matrix $N$, we form a new tree $T'$ by attaching a path $P$ of length $a$ at $s_{N}$, and letting $s'$ be the other endpoint of $P$. Then $(T',s', V_N)$ is a base of $\phi_a(N)$.
	
	\item Extension II (with parameters $a\geq b>0$): Given a base $(T_N, s_{N}, V_{N})$ corresponding to the Gromov matrix $N$, we form a new tree $T'$ by attaching a path $P$ of length $a$ at $s_{N}$ as in extension I, with $s'$ being the other endpoint of $P$. 
	The path $P$ is constructed to have a vertex $v$ whose distance to $s'$ is $b$ (it is possible that $v=s_N$). Then, the base $(T',s',V_N\cup \{v\})$ corresponds to $\phi_{a,b}(N)$.
\end{itemize}  

\begin{figure}[!htb] 
	\centering
	\includegraphics[width=0.65\textwidth]{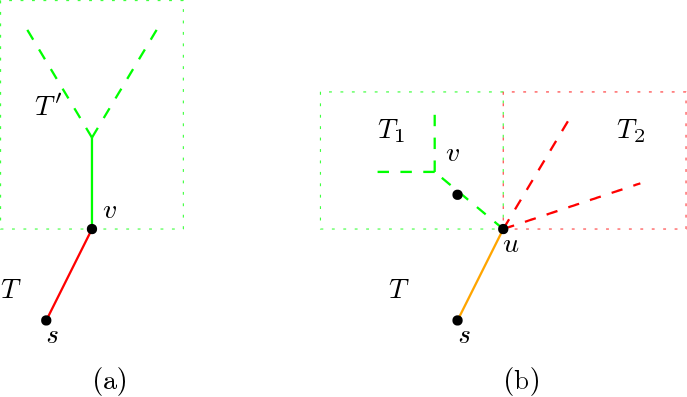}
	\caption{Illustration of the two cases in the induction step.} \label{fig:8}
\end{figure}

We now proceed to prove Proposition~\ref{prop:mit} for the base $(T,s,V)$ by induction on $n=|V|$. As we assume that $T$ is non-empty, $V$ is non-empty. The case $|V|=1$ requires an initialization step. Suppose that the proposition is true when the base set has less than $n$ vertices. Let $v\in V$ be a node such that $[v,s]\cap V=\{v\}.$ There are two cases to consider (see \figref{fig:8} for an illustration):
\begin{enumerate}[(a)]
	\item The path $P= (v,s]$ does not contain any vertex of degree greater than $2$. Let $T'=T\backslash P$, $s_{T'}=v$ and $V_{T'}=V_T\backslash \{v\}$. As $|V_{T'}|=|V_T|-1$, by the induction hypothesis, we can obtain $(T', s_{T'}, V_{T'})$ by the Gromovication operations. Moreover, $(T, s, V)$ can be constructed from $(T', s_{T'}, V_{T'})$ by an extension II.
	\item The path $P= (v,s)$ contains of a vertex $u$ of degree greater than $2$. By our choice of $v$, the node $u\notin V$. We can always find two connected subtrees $T_1$ and $T_2$ of $T\backslash P$ such that:
	\begin{enumerate}[(i)]
		\item $T_1\cap T_2 =u$;
		\item $T_1\cup T_2 = T\backslash P$;
		\item $T_i \cap V \neq \emptyset$, for $i=1,2$.
	\end{enumerate}
	Let $s_{T_1}=s_{T_2}=u$; and $V_{T_i}=T_i \cap V$, for $i=1,2.$ The induction hypothesis implies that both $(T_i, s_{T_i}, V_{T_i}), i=1,2$ can be obtained from Gromovication operations. Moreover, $(T, s, V)$ can be obtained from $(T_i, s_{T_i}, V_{T_i})$, $i=1,2$ by a direct sum followed by possibly an extension I.
\end{enumerate} 
The proof is now complete.

\section{Proof of Proposition~\ref{prop:wht}}\label[appendix]{proof:prop:wht}

\begin{enumerate}[(a)]
	\item This is obvious.
	\item 
	Recall that a special case of Weyl's inequality (cf.\cite{Fra93}) asserts the following: For two symmetric matrices $M$ and $N$ of the same size, the inequality 
	\begin{align}\label{ineq:Weyl}
	\lambda_{\min}(M+N)\geq \lambda_{\min}(M)+\lambda_{\min}(N)
	\end{align}
	holds. In view of Weyl's inequality, it suffices to note that the $n\times n$ matrix with identical entries $a$ has two eigenvalues $na$ and $0$.
	
	\item Let ${\bf 1}_n$ be the $n\times n$ square matrix with all entries $1$. If $a>b$, we form two $(n+1)\times (n+1)$ matrices: $M_1 = M\oplus (b-b^2/a)$; and $M_2$ has the $n\times n$ block $a{\bf 1}_n$ as its top-left submatrix, $b^2/a$ in the bottom-right corner, and the remaining entries $b$. The column rank of $M_2$ is $1$ and it is easy to check that $\lambda_{\min}(M_2)=0.$ Moreover, from item~\ref{prop:wht:a}, $\lambda_{\min}(M_1) = \min\{\lambda_{\min}(M), b-b^2/a\}$. Therefore, as $\phi_{a,b}(M)=M_1+M_2$, we have the desired inequality $$\lambda_{\min}(\phi_{a,b}(M))\geq \min\{\lambda_{\min}(M),b-b^2/a\}$$ by Weyl's inequality~\ref{ineq:Weyl}.
	
	Suppose $a=b$. Let $\alpha=\lambda_{\min}(M)/a$. We form two $(n+1)\times (n+1)$ matrices: $M_1 = M\oplus (a\alpha)$; and $M_2$, which has $a(1-\alpha)$ in the bottom-right corner and $a$ everywhere else. As $\phi_{a,a}(M)=M_1+M_2$, by Lemma \ref{lem:fea} below and Weyl's inequality, we have 
	\begin{equation*}
	\begin{split}
	& \lambda_{\min}(\phi_{a,a}(M))\\
	& \geq \lambda_{\min}(M_1)+\lambda_{\min}(M_2)\\
	& = a\alpha + a\frac{n+1-\alpha-\sqrt{(n+1-\alpha)^2+4n\alpha}}{2} \\
	& = a\cdot \frac{n+1+\alpha-\sqrt{(n+1+\alpha)^2-4\alpha}}{2} \\
	& \geq \frac{a\alpha}{n+1+\alpha} = \frac{\lambda_{\min}(M)}{n+1+\lambda_{\min}(M)/a}.
	\end{split}
	\end{equation*}		
\end{enumerate}
The proof of the proposition is now complete.

\begin{Lemma_A} \label{lem:fea}
	Suppose that $M$ is the $n\times n$ matrix whose bottom-right corner is $1-\alpha$ for some $\alpha >0$, and has $1$ everywhere else. Then 
	\begin{align*}
	\lambda_{\min}(M) = \frac{n-\alpha-\sqrt{(n-\alpha)^2+4(n-1)\alpha}}{2}.
	\end{align*} 
\end{Lemma_A}
\begin{IEEEproof}
	It is enough to show that the characteristic polynomial of $M$ is $(-1)^n\lambda^{n-2}(\lambda^2-(n-\alpha)\lambda-(n-1)\alpha).$ We give a sketch of the proof with some calculation details omitted. First we note that $M$ has $n-1$ identical columns, the characteristic polynomial takes the form $(-1)^n\lambda^{n-2}(\lambda^2-a\lambda -b)$, where the coefficient $a$ is the trace $n-\alpha$. 
	
	Let $M'=M-\lambda I_n$, where $I_n$ is the $n\times n$ identity matrix. To determine $b$, let $C(i,j)$ be the $(i,j)$-th cofactor of $M'$; and $M'(i,j)$ the $(i,j)$-th entry of $M'$. The determinant $\det(M')=\sum_{1\leq j\leq n} M'(1,j)C(1,j).$ The summand $M'(1,1)C(1,1)= (1-\lambda)C(1,1)$ can be computed by induction as $$(1-\lambda)(-1)^{n-1}\lambda^{n-3}(\lambda^2-(n-1-\alpha)\lambda-(n-2)\alpha).$$ For the remaining summands, only the leading coefficient of $\lambda^{n-2}$ contributes to $b$, which is either $1$ or $-1$; and the coefficient of $\lambda^{n-2}$ of $\sum_{1< j\leq n} M'(1,j)C(1,j)$ is $(-1)^{n-1}(n-1)$. It is now easy to obtain that $b=(n-1)\alpha.$
\end{IEEEproof}

\section{Proof of Theorem~\ref{thm:fav}}\label[appendix]{proof:thm:fav}

We first prove an elementary lemma.

\begin{Lemma_A} \label{lem:ftr}
	For a triplet of real numbers $A=\{a_1, a_2, a_3\}$, the new triplet $A'=\{a_1', a_2', a_3'\}$ defined by $a_i'=\max\{a_i,\min(A\backslash \{a_i\})\}$ satisfies the following properties:
	\begin{enumerate} [(a)]
		\item $|A\cap A'|\geq 2;$
		\item $\max(A) = \max(A');$
		\item The triplet $A'$ satisfies the three-point condition. 
	\end{enumerate}
\end{Lemma_A}
\begin{IEEEproof}
	Without loss of generality, we assume that $a_1\geq a_2\geq a_3$. It is easy to verify that $a_1'=a_1, a_2'=a_2$ and $a_3'=a_2.$ Therefore, the lemma follows.
\end{IEEEproof}

For convenience, we call $a_i'$ the \emph{max-min} of $a_i$ w.r.t.\ $A$. Suppose we have a pair of functions $a(\alpha)$ and $b(\alpha)$. If both of them are piecewise linear continuous in $\alpha$, then so are $\max(a,b)$ and $\min(a,b)$. Similarly, suppose we have functions $a_1(\alpha), a_2(\alpha), a_3(\alpha)$ such that all of them are piecewise linear continuous in $\alpha$. Then the same holds for the max-min of $a_i(\alpha),i=1,2,3$, w.r.t.\ $\{a_1(\alpha),a_2(\alpha),a_3(\alpha)\}$. 

We now proceed with the proof of Theorem~\ref{thm:fav}. We have the following alternative (though inefficient) way of obtaining $M_{\alpha}'$ from the convex combination $M_{\alpha}$. The matrices constructed below depends on $\alpha$, but for convenience, we suppress $\alpha$ in the notations. We start with $N_0 = M_{\alpha}$ and $N_i$ is obtained from $N_{i-1}$ as follows: given $j<k$ and $l\neq j,k$, we first form the triple $A_{j,k,l} = \{N_{i-1}(j,k), N_{i-1}(l,j), N_{i-1}(l,k)\}$. Next, we find $a_{j,k}^{l}$ as the max-min of $N_{i-1}(j,k)$ w.r.t.\ $A_{j,k,l}$. The $(j,k)$-th entry of $N_i$ is taken as $\max_{l\ne j,k} a_{j,k}^{l}$. During this process, the only functions involved are $\max$ and $\min$. Therefore, if the entries of $N_{l-1}$ are piecewise linear in the parameter $\alpha$, so are the entries of $N_l$. Hence $N_l$ itself is piecewise linear in $\alpha$ when considered as an element of the vector space of $n\times n$ square matrices. 

To conclude the proof of the theorem, we need to show two additional assertions:
\begin{enumerate}[(i)]
	\item\label{it:iterations} The sequence of matrices $N_i, i\geq n$ stabilizes (the entries do not change in the next iteration) after a fixed number of iterations.
	\item\label{it:stabilized} The stabilized matrix is $M_{\alpha}'$.
\end{enumerate}

For \ref{it:iterations}, we make the following observation: the largest off-diagonal entries of $N_0$ do not change in subsequent iterations. Denote by $S_i$ the positions of the entries of $N_{i}$ that do not change in all subsequent iterations, which is non-empty by the case $i=0$. On the other hand, the largest off-diagonal entries of $N_{i+1}$ with positions different from $S_i$ do not change in all subsequent iterations. Therefore $|S_{i+1}|>|S_i|$ unless $N_{i+1}=N_i$. Therefore the sequence of matrices stabilizes within $n(n-1)/2$ iterations. 

For \ref{it:stabilized}, it is easy to see that the construction given in Section \ref{sec:con} is merely an economized alternative for $M_{\alpha}'$. The result thus follows.

\section{Some Geometric Constructions}\label[appendix]{geometric}

Suppose $M$ is a Gromov matrix with base $(T,s,V)$, where $V=\{v_1,\ldots, v_n\}$. For each vertex pair $v_i,v_j$, the distance $d_T(v_i,v_j)$ between them can be recovered as $M(i,i)+M(j,j)-2M(i,j).$ The nodes in $V$ uniquely determine $T$ as $V\cup \{s\}$ span $T$. However, we are not able to recover the (weighted) adjacency matrix of $T$ directly using such information, as $V$ may not contain all the vertices of $T$ in the usual sense (for a simple example, see \figref{fig:1}). On the other hand, in certain applications, we are interested in the relative positions of vertices of $V$ on $T$, instead of the relative distance (e.g., some applications may require only knowledge of the number of neighbors of certain vertices, or the size of the connected components after removing a particular node). In this appendix, we discuss certain geometric constructions based on $(T,s,V)$, or equivalently $M$, which allows us to explore the relative positions of the vertices of $V$.

We first introduce a graph $G_{V}$ as follows: the size of $G_{V}$ is $|V|=n$ and let $U=\{u_1,\ldots, u_n\}$ be the vertex set of $G_V$. Two vertices $u_i$ and $u_j$ are connected by an edge in $G_V$ if and only if $[v_i,v_j]\cap V=\{v_i,v_j\}$ in $T$. The weight of each edge of $G_V$ is set to be $1$, and the distance metric on $G_V$ is denoted by $d_{G_V}(\cdot,\cdot)$.

\begin{figure}[!t] 
	\centering
	\includegraphics[width=0.55\textwidth]{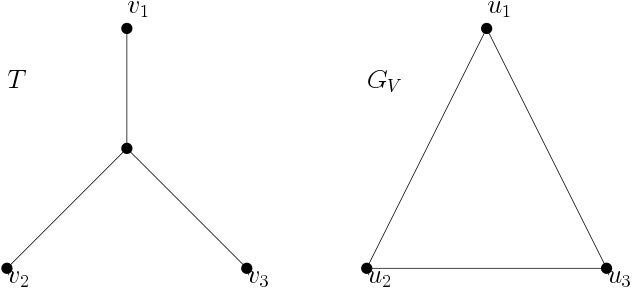}
	\caption{In tree $T$, consider $V=\{v_1,v_2,v_3\}$. The associated $G_V$ (shown on the right) is no longer a tree.} \label{fig:6}
\end{figure}

For each path $P$ (not required to be simple) in $G_V$, we form an associated path $P'$ in $T$ as follows: if $P$ is obtained from concatenating edges $(u_i,u_j)$, we construct $P'$ by concatenating simple paths $[v_i,v_j]$. If $P$ with length $l$ is a path connecting $u_i$ and $u_j$ such that $d_{G_V}(u_i,u_j)=l$, then we call $P$ a \emph{geodesic} path for convenience. One should take note here that $G_V$ is not a tree in general (cf.\ \figref{fig:6}), and there can be multiple geodesic paths between two nodes.

\begin{figure}[!t] 
	\centering
	\includegraphics[width=0.45\textwidth]{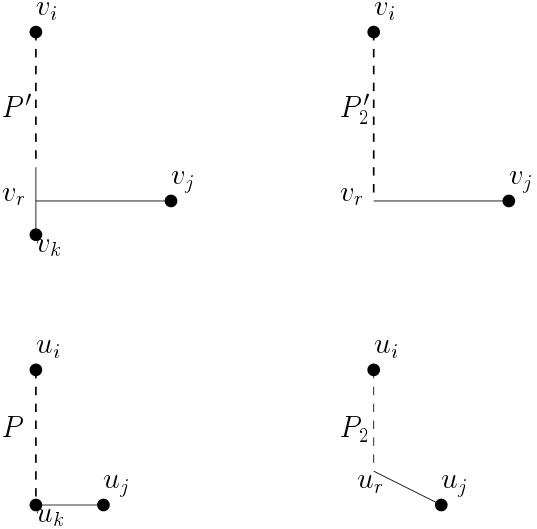}
	\caption{Illustration of the case $[v_k,v_j]\cap P_1' \neq \{v_k\}$ in the proof of Lemma~\ref{lem:ifpi}.} \label{fig:7}
\end{figure}

\begin{Lemma_A} \label{lem:ifpi}
	If $P$ is a geodesic path connecting $u_i$ and $u_j$, then the associated path $P'$ in $T$ is simple.	
\end{Lemma_A}
\begin{IEEEproof}
	Let the length of $P$ be $l$. We prove the claim by induction on $l$. The case $l=1$ clearly holds by definition.
	
	Suppose the claim holds for $l=t$. Consider $l=t+1$. Let $P_1=[u_i,u_k]$ be the subpath of $P$ starting from $u_i$ of length $t$. By the induction hypothesis, the associated path $P_1'$ in $T$ is simple, connecting $v_i$ and $v_k$. If the intersection $[v_k,v_j]\cap P_1' = \{v_k\}$ in $T$, then $P'$ as the concatenation of $P_1'$ and $[v_k,v_j]$ is simple; and this proves the lemma.
	
	Otherwise, $[v_k,v_j]\cap P_1' \neq \{v_k\}$ (see \figref{fig:7} for an illustration). In this case, there exists a $v_r\ne v_k$ in $P'$ and $v_r\in[v_k,v_j]\cap P_1'$. We can form another path $P_2$ consisting of $[u_i,u_r]$ concatenated with $[u_r,u_j]$ in $G_V$. The path $P_2$ is of length strictly smaller than $l$; and hence $d_{G_V}(u_i,u_j)<l$. This contradicts the assumption that $P$ is a geodesic path.
\end{IEEEproof}

Since $G_V$ may not be a tree (cf.\ \figref{fig:6}), and we have the following weaker substitute.

\begin{Corollary_A}
	For each $u\in U$, there is a unique BFS tree of $G_V$ based at $u$.	
\end{Corollary_A}

\begin{IEEEproof}
	Suppose on the contrary that for $u_i\in U$ there are two distinct BFS trees $T_1$ and $T_2$ of $G_V$. This means that there is a vertex $u_j\neq u_i\in U$ such that the following holds:
	\begin{enumerate}[(a)]
		\item The path $[u_i,u_j]$ in $T_1$ and $T_2$ are distinct; and denote them by $P_1$ and $P_2$ respectively.
		\item $P_1$ and $P_2$ are of the same length $d_{G_V}(u_i,u_j)$.	
	\end{enumerate}
	
	Let $P_1'$ and $P_2'$ be their associated paths (connecting say $v_i$ and $v_j$) in $T$ respectively. As $P_1$ and $P_2$ are distinct, so are $P_1'$ and $P_2'$. Moreover, by Lemma \ref{lem:ifpi}, both $P_1'$ and $P_2'$ are simple paths. This is absurd, as $T$ is a tree and there is a unique simple path connecting $v_i$ and $v_j$. This concludes the proof of the corollary by contradiction.
\end{IEEEproof}

From this result, we see that $G_V$ resembles a tree in the sense that each node has a unique BFS tree. It is an important feature to handle information propagation. On the other hand, $U$ contains all the nodes of $G_V$. Therefore, we may use $G_V$ to study the relative positions of nodes in $V$ of $T$, as well as other combinatorial properties. 

Now we indicate how to obtain the adjacency matrix $A$ of $G_V$ from the Gromov matrix $M$. To determine the $(i,j)$-th entry $A(i,j)$ of $A$, consider $v_i$ and $v_j$ in $T$. As we have demonstrated, the distance $d_T(v_i,v_j)$ can be recovered from $M$. Moreover, recall that $T$ is a tree. Therefore $A(i,j)=1$ if and only if for each $1\leq k\neq i,j \leq n$, $d_T(v_i,v_j)\neq d_T(v_i,v_k)+d_T(v_j,v_k)$; or equivalently, 
$$M(k,k)+M(i,j)\neq M(i,k)+M(k,j).$$ Reconstructing $A$ thus requires a complexity of $O(n^3)$.

\bibliographystyle{IEEEtran}
\bibliography{IEEEabrv,StringDefinitions,all,CDN}

\end{document}